\documentclass[11pt]{article}

\usepackage[a4paper, margin=1in]{geometry}

\usepackage{amsthm,amsmath,amsfonts,amssymb}
\usepackage{microtype}

\usepackage[authoryear, round]{natbib}

\usepackage[colorlinks, citecolor=blue, urlcolor=blue, linkcolor=blue]{hyperref}
\usepackage{xurl}

\usepackage{parskip}

\usepackage{graphicx}
\graphicspath{{./images/}}
\usepackage{float}
\usepackage{subcaption}
\usepackage{booktabs}

\usepackage[linesnumbered, ruled, vlined]{algorithm2e}

\usepackage{tikz}
\usetikzlibrary{shapes,arrows,fit,calc,positioning}
\tikzset{box/.style={draw, diamond, thick, text centered, minimum height=0.5cm, minimum width=1cm}}
\tikzset{line/.style={draw, thick, -latex'}}

\usepackage{comment}
\usepackage{dsfont}

\theoremstyle{plain}

\theoremstyle{definition}

\theoremstyle{remark}

\newcommand{\E}{\mathbb{E}}

\newcommand{\p}{\mathbb{P}}
\renewcommand{\H}{\mathcal{H}}
\newcommand{\T}{\mathcal{T}}
\newcommand{\ind}{\perp\!\!\!\perp}

\makeatletter
\newcommand{\fakeappendixlabels}{
  \newlabel{appendix:outer_gibbs}{{S.1}{0}}
  \newlabel{appendix:comp_details}{{S.2}{0}}
  \newlabel{appendix:main_sim}{{S.3}{0}}
  \newlabel{appendix:additional_sim}{{S.4}{0}}
  \newlabel{appendix:clinical_covariates}{{S.5}{0}}
  \newlabel{appendix:landmark}{{S.6}{0}}
}
\makeatother
\fakeappendixlabels

\title{Horseshoe Forests for High-Dimensional Causal Survival Analysis}

\author{
  Tijn Jacobs\thanks{Department of Mathematics, Vrije Universiteit Amsterdam. \texttt{t.jacobs@vu.nl}} \and
  Wessel N.\ van Wieringen\thanks{Department of Epidemiology and Data Science, Amsterdam University Medical Centre. \texttt{w.n.van.wieringen@vu.nl}} \and
  St\'ephanie L.\ van der Pas\thanks{Department of Mathematics, Vrije Universiteit Amsterdam. \texttt{s.l.vander.pas@vu.nl}}
}

\date{\today}

\begin{document}

\maketitle

\begin{abstract}
We develop a Bayesian tree ensemble model to estimate heterogeneous treatment effects in censored survival data with high-dimensional covariates.
Instead of imposing sparsity through the tree structure, we place a horseshoe prior directly on the step heights to achieve adaptive global--local shrinkage.
This strategy allows flexible regularisation and reduces noise.
We develop a reversible jump Gibbs sampler to accommodate the non-conjugate horseshoe prior within the tree ensemble framework.
We show through extensive simulations that the method accurately estimates treatment effects in high-dimensional covariate spaces, at various sparsity levels, and under non-linear treatment effect functions.
We further illustrate the practical utility of the proposed approach by a re-analysis of pancreatic ductal adenocarcinoma (PDAC) survival data from The Cancer Genome Atlas.

\medskip
\noindent\textbf{Keywords:} Causal inference; heterogeneous treatment effects; survival analysis; high-dimensional; tree ensembles; shrinkage priors.
\end{abstract}

\section{Introduction}

Modelling heterogeneous treatment effects is challenging, especially with survival outcomes and high-dimensional covariates that may confound or modify treatment effects.
We propose a Bayesian regression tree ensemble approach that uses a horseshoe prior to shrink tree step heights.
This framework flexibly captures complex non-linear effects and interactions while providing adaptive shrinkage suited to high-dimensional causal inference with survival outcomes.


The conditional average treatment effect (CATE) quantifies how treatment effects vary across covariate profiles and is a central object in modelling heterogeneous treatment effects \citep{Caron2022CATE}.
Estimating the CATE requires careful adjustment for confounding variables.
This task becomes particularly challenging in high-dimensional settings with many covariates and complex interactions. 
In contrast, the average treatment effect (ATE) is low-dimensional and has been widely studied in high-dimensional contexts \citep{Farrell2015RobustATE, Antonelli2018DRMatching, Antonelli2020Averaging, Ning2020HDPS, Antonelli2022HighDimCausal}.
Researchers can treat high-dimensional components as nuisance terms when estimating the ATE, which simplifies inference.
CATE estimation is fundamentally more difficult because the parameter of interest is high-dimensional and depends directly on covariates \citep{Alaa2018LimitsHTE}.
Recent work has proposed flexible and robust methods for estimating heterogeneous treatment effects in high-dimensional covariate spaces \citep{Powers2018HTE, Fan2020CATE, Shin2023DRInference}.
Despite this progress, a gap remains in extending these high-dimensional causal inference methods to survival data, where censoring and time-to-event outcomes introduce additional complexity.

Tree-based methods are widely used for causal inference in both frequentist and Bayesian frameworks. 
On the frequentist side, the causal forest algorithm of \citet{Wager2018CausalForest} extends the sample-splitting approach of \citet{Athey2016CausalTrees} to estimate heterogeneous treatment effects non-parametrically. 
This approach has been adapted to survival data by \citet{Cui2023CausalSurvivalForest}. 
In the Bayesian setting, the Bayesian causal forest approach of \citet{hahn2020bayesian} extends earlier work by \citet{Hill2011}. 
Hill applied Bayesian additive regression trees (BART) \citep{BARToriginal} to estimate treatment effect heterogeneity.
BART-based methods have demonstrated strong empirical performance and often outperform alternative estimators in benchmarking studies \citep{Dorie2019CausalCompetition, Thal2023ACIC}. 
Researchers have adapted these models to survival data to estimate conditional average treatment effects \citep{Bonato2011BayesianEnsemble, henderson2020individualized, Hu2023FlexibleCausal, Sun2025AFTCure}.
Recent work shows that AFT-BART \citep{henderson2020individualized} outperforms other causal machine learning approaches in survival settings \citep{Hu2021HeterogeneousSurvival}.

We adapt Bayesian additive regression trees to high-dimensional data by placing a global--local shrinkage prior on the step heights.
In the original BART formulation, regularisation is imposed primarily through constraints on the tree structure, such as limiting tree depth or controlling the likelihood of node splits \citep{BARToriginal}.
More recent extensions, such as the Dirichlet additive regression trees (DART) of \citet{linero2018dirichlet}, introduced Dirichlet priors on splitting proportions to encourage sparsity in the covariate space.
\citet{caron2022shrinkage} further extended this idea to causal inference. 
Our approach departs from this line of work by shifting the focus of regularisation away from the tree structure and onto the leaf node parameters.
This approach preserves the flexibility of BART to model complex non-linear relationships. 
At the same time, it introduces adaptive shrinkage that suits high-dimensional causal inference. 
Rather than enforcing exact sparsity, our method applies continuous shrinkage that down-weights weak effects while still allowing strong signals to be detected when supported by the data.

Causal inference typically relies on the \textit{unconfoundedness} assumption to identify treatment effects from observational data \citep{ImbensRubin2015}.
It requires that, conditional on observed covariates $X$, treatment assignment $A$ is independent of the potential outcomes $T(a)$:
\begin{equation}\label{unconfoundedness}
T(a) \ind A \mid X = x \quad \text{for } a \in \{0, 1\}.
\end{equation}
Valid causal inference is possible by adjusting for the relevant covariates if this condition holds.
Violating unconfoundedness may lead to biased estimates \citep{Chernozhukov2022OVBCML}.

In high-dimensional settings such adjustment is inherently affected by regularisation, as the estimand itself is a function of $X$.
Although treatment effect heterogeneity may be complex, it is typically assumed to be driven by an unknown low-dimensional structure in the covariates, such as a small subset of effect modifiers and confounders.
When $p$ is large relative to $n$, effective adjustment may benefit from regularisation to control model
complexity and to distinguish signal from noise.

\cite{hahn2018regularization} show that regularisation can induce bias even when the outcome model is correctly specified and all confounders are observed. 
This phenomenon, termed \textit{regularisation-induced confounding} (RIC), arises when shrinkage attenuates adjustment for covariates associated with treatment assignment, thereby distorting the estimated treatment effect. 
The risk in high-dimensional settings is particularly pronounced for weak confounders that strongly predict treatment but are only weakly associated with the outcome.
We adopt a modelling strategy that retains all covariates and controls complexity through adaptive shrinkage rather than variable exclusion.
We investigate the resulting bias empirically in simulation settings designed to exhibit RIC \citep{hahn2020bayesian}.
Motivated by this interaction between regularisation and confounding, we employ global--local shrinkage priors for adaptive regularisation of treatment effect heterogeneity.

Global-local shrinkage priors provide a fully Bayesian and computationally attractive alternative to exact sparsity-inducing methods \citep{Mitchell1988SpikeSlab, Liu2021ABC}. 
Originally developed for sparse regression problems \citep{OriginalHorseshoe}, these priors are now used across a wide range of contexts \citep{Kowal2019DynamicShrinkage, Li2019GraphicalHorseshoe, Louizos2017BayesianCompression} and admit many formulations \citep{tipping2001, johnstone2004, griffin2005, Bhadra2019}\citep{park2008}. 
Among these, the horseshoe prior has emerged as a particularly attractive choice because of its strong theoretical properties and empirical performance in high-dimensional settings \citep{vdPas2017Horseshoe, hahn2020bayesian, DattaGhosh2013, vdPas2014, vdPas2017, bhadra2019lasso}. 
We focus on the horseshoe prior to introduce continuous shrinkage directly on the tree step heights.
This approach retains all covariates while adaptively shrinking their contributions. 
The proposed methodology is readily extendable to other global-local shrinkage priors that allow Gibbs sampling and could be further generalised to incorporate different shrinkage strategies with additional computational effort.

Our work contributes to the literature in several ways:
\begin{enumerate}
    \item We introduce a novel regularisation strategy for Bayesian Additive Regression Trees based on the step-height prior, where we use the horseshoe prior to induce continuous shrinkage rather than relying solely on regularisation via the tree structure.
    \item We develop a computational framework for fitting these models via reversible jump Metropolis-Hastings. The implementation is publicly available in the \texttt{R} package \texttt{ShrinkageTrees} \citep{ShrinkageTreesR}.
    \item We demonstrate how this framework can be effectively applied to estimate heterogeneous treatment effects in survival data, with a particular focus on high-dimensional covariates.
\end{enumerate}

The remainder of this paper is organised as follows.
In Section~\ref{sec:context} we describe the causal framework and notation.
In Section~\ref{section:Model} we introduce the model formulation and prior specification. 
In Section~\ref{sec:sampling} we describe the computational strategy and posterior inference procedure. 
We present the simulation results in Section~\ref{section:Simulations}.
Section~\ref{section:PDAC} provides a detailed case study of pancreatic cancer data. 
We conclude with a discussion in Section~\ref{section:Discussion}.

\section{Causal framework}\label{sec:context}

\subsection{Modelling of causal effects}

Let $T$ denote the non-negative survival time and $C$ denote the censoring time. 
We observe the follow-up time $Y = \min(T, C)$ along with the censoring indicator $\delta \in \{0,1\}$, where $\delta = 1$ if the event is observed and $\delta = 0$ if the observation is right-censored.
Let $A \in \{0,1\}$ be a binary treatment indicator, where $A = 1$ corresponds to the treated group and $A = 0$ to the control group. 
Covariates are denoted by $X \in \mathbb{R}^p$, where $p$ is the number of observed pre-treatment variables.
The covariates may include potential confounders.
We do not distinguish between various types of covariates in our notation, such as confounders, prognostic factors, moderators, or predictors of treatment assignment \citep{herren2020fast}.
Throughout, we use capital letters (e.g., $T$, $A$, $X$) to denote random variables, and lowercase letters (e.g., $t$, $a$, $x$) to denote their realisations. 
We consider a sample of $n$ independent observations. 

We assume an accelerated failure time (AFT) model \citep{buckley1979linear} for the potential survival times. 
The AFT model relates the logarithm of the survival time to a regression function.
We decompose the log-transformed outcome into a \textit{prognostic} component and a \textit{treatment effect} component \citep{hahn2020bayesian}:
\begin{equation}\label{model}
    \log T(a) = f(x, \hat{e}(x)) + a \cdot \tau(x) + \varepsilon,
\end{equation}
where $\varepsilon \sim \mathcal{N}(0, \sigma^2)$ denotes the error term. 
The function $f(x, \hat{e}(x))$ models the baseline prognosis and incorporates the estimated propensity score $\hat{e}(x)$.
The propensity score, $e(x) = \mathbb{P}(A = 1 \mid X = x)$, is the probability of treatment given covariates \citep{RosenbaumRubin1983}.
The function $\tau(x)$ directly models the treatment effect.
In Section~\ref{section:Model}, we describe a novel regression framework for $f$ and $\tau$ that builds on Bayesian regression trees and incorporates a horseshoe prior to handle the high-dimensional covariates.
We refer to model \eqref{model} with horseshoe prior on the step heights of $f$ and $\tau$ as the \textsl{Causal Horseshoe Forest}.

The AFT model offers several advantages in the context of causal inference with survival data. 
The AFT model provides a \textit{collapsible} estimand: the acceleration factor, defined as the ratio of expected survival times under treatment and control, remains invariant when conditioning on additional non-confounding covariates \citep{Robins1992AFT, Aalen2015CoxCausal, Crowther2023}.
This implies that $\tau(x)$ can be meaningfully averaged across covariate distributions to obtain the marginal causal effect without introducing bias due to non-collapsibility.
Collapsibility is a desirable property in causal inference, as it ensures interpretability and comparability of marginal and conditional effects. 
In contrast, commonly used hazard-based measures, such as the hazard ratio in the Cox model, are non-collapsible \citep{Hernan2010, Martinussen2013, Aalen2015CoxCausal}.
Furthermore, \cite{Hougaard1999} showed that AFT models are robust to omitted covariates, enhancing their suitability for observational studies.

We include the estimated propensity score in the outcome model to improve causal effect estimation in high-dimensional settings. 
This inclusion helps to control for confounding and mitigates regularisation-induced confounding, where aggressive shrinkage of confounders can bias treatment effect estimates \citep{hahn2020bayesian, Ray2020SemiparametricBayes}. 
The propensity score acts as a balancing score: given the propensity score, covariates and treatment assignment become independent \citep{RosenbaumRubin1983}. 
We incorporate the \textit{estimated} propensity score directly in the prognostic regression function $f(x, \hat{e}(x))$ to avoid feedback between the outcome and treatment models.
Although this approach is not dogmatically Bayesian and does not propagate uncertainty from the propensity score estimation \citep{Zigler2013ModelFeedback}, it improves performance in finite samples \citep{hahn2020bayesian}. 
We estimate the propensity score using a tree-based model specifically designed for high-dimensional covariates and binary treatment assignment.

\subsection{Estimands and identification assumptions}

We define causal estimands within the potential outcomes framework \citep{Rubin1974, Rubin1978, Neyman1990}. 
For each individual, let $T(1)$ and $T(0)$ represent the survival times that would be observed under treatment and control, respectively. 
Likewise, $C(1)$ and $C(0)$ denote the potential censoring times corresponding to each level of treatment.
We focus on the conditional average treatment effect (CATE), defined as the difference in expected log survival times under treatment and control, conditional on observed covariates:
\begin{equation}
    \text{CATE}(x) := \E[\log T(1) - \log T(0) \mid X = x].
\end{equation}
The CATE captures heterogeneity in treatment effects across subgroups defined by covariate profiles $X = x$.  
We emphasise that the CATE is a population-level functional of the conditional potential outcome distributions.
It should be interpreted as a function of covariates $x$ and not as a realised individual-level causal contrast, since it depends only on the marginal conditional expectations $\E[\log T(a)\mid X=x]$ for $a\in\{0,1\}$ and does not involve the joint distribution of potential outcomes.

We rely on common assumptions to identify $\tau(x)$ from observational data \citep{ImbensRubin2015}.
The first is the \textit{Stable Unit Treatment Value Assumption} (SUTVA), which rules out interference across units and presumes each individual has well-defined potential outcomes under both treatment conditions.
This implies that the observed survival time can be expressed as $T = A T(1) + (1 - A) T(0)$.
We assume \textit{unconfoundedness} (see \eqref{unconfoundedness}) and \emph{positivity}.
Positivity requires that for all covariate profiles $x $ each treatment arm has a non-zero probability of assignment:
\begin{equation}
0 < e(x)  < 1.
\end{equation}
The assumptions above are sufficient to identify the causal estimands in terms of the survival times $T$ conditional on $(X,A)$. 
An additional assumption on the censoring mechanism is required to identify these quantities from the observed, right-censored data.
We assume \emph{independent censoring}, meaning that the censoring mechanism is conditionally independent of the survival time given treatment and covariates:
\begin{equation}
C(a) \ind T(a) \mid X = x, A=a \quad\text{for } a \in \{0,1\}.
\end{equation}
Under this assumption, the observed data $(Y,\delta)$ are sufficient to identify the conditional distribution of $T$ given $(X,A)$ from right-censored observations \citep{Andersen1993, Lagakos1979}.
Conditional on $(X,A)$, the censoring mechanism does not depend on the event time.
The observed events are representative of the underlying event-time distribution within each covariate--treatment stratum.
As a result, functionals of the conditional distribution of $T$, including $\E[\log T \mid X=x,A=a]$, can be identified from the censored data.

The causal effect is identifiable from the observed data under these assumptions \citep{Imbens2004Identification}, i.e.  \protect{$\E[\log T(a) \mid X = x] = \E[\log T \mid X = x, A = a]$} for each $a \in \{0,1\}$.
As a result, the CATE can be expressed in terms of observable quantities as:
\begin{equation}
\begin{split}
    \text{CATE}(x) &= \E[\log T \mid X = x, A = 1] - \E[\log T \mid X = x, A = 0] =\tau(x).
\end{split}
\end{equation}
The final equality follows from model assumption \eqref{model}. 
This identification enables a direct modelling approach of the conditional average treatment effect.
\section{Bayesian regression trees}\label{section:Model}

In this section, we describe how we adapt the Bayesian Additive Regression Trees (BART) prior \citep{BARToriginal} to high-dimensional covariates by introducing a horseshoe prior on the step heights.
This approach shifts regularisation from the tree structure (e.g. \citet{linero2018dirichlet}) to the step height parameters.
We start with an overview of how BART is applied to model log survival times under the accelerated failure time (AFT) framework.
We detail our novel adaptation using a global-local shrinkage prior.
We then explain how the model accommodates censored outcomes.
Finally, we discuss default hyperparameter choices.

\subsection{Overview of Bayesian Additive Regression Trees (BART)}

We adapt the BART model to estimate the conditional mean function $\mathbb{E}[\log{T_i}\mid X_i = x]$ in the AFT model.
The uncensored survival times are modelled as a noisy sum of trees:
\begin{equation}\label{bart_model}
\log{T_i} = \sum_{j=1}^m g(X_i; \mathcal{T}_j, \mathcal{H}_j) + \varepsilon_i, \quad \varepsilon_i \sim \mathcal{N}(0, \sigma^2),
\end{equation}
where each $\mathcal{T}_j$ is a binary decision tree that defines recursive partitioning rules over the covariate space, and $\mathcal{H}_j = \{h_{j1}, \ldots, h_{jL_j}\}$ denotes the set of step heights associated with the $L_j$ leaves of tree $\mathcal{T}_j$.
We extend this model to accommodate censored survival times in Section~\ref{cens_bin}.

The function $g(x; \mathcal{T}_j, \mathcal{H}_j)$ maps $x$ to a leaf node $\ell$ in $\mathcal{T}_j$ and returns the corresponding value $h_{j\ell}$. 
Each tree partitions the covariate space into disjoint regions, assigning a constant prediction to each region (see Figure~\ref{ExampleTree}).

\begin{figure}[tb]
    \centering
    \begin{minipage}{.5\linewidth}
      \centering
      \begin{tikzpicture}[
            box/.style={rectangle, draw, rounded corners, align=center, minimum height=10mm, minimum width=16mm},
            leaf/.style={circle, draw, align=center, minimum size=10mm},
            line/.style={-latex}
        ]
        
        \node [box] (root) {$x_1 < 0.7$};
    
        \node [box, below=1cm of root, xshift=-2cm] (left) {$x_2 < 0.6$};
        \node [leaf, below=1cm of root, xshift=2cm] (h1) {$h_1$};
    
        \node [leaf, below=1cm of left, xshift=-1cm] (h2) {$h_2$};
        \node [leaf, below=1cm of left, xshift=1cm] (h3) {$h_3$};
    
        \draw[line] (root) -- (left) node[midway, above left] {Yes};
        \draw[line] (root) -- (h1) node[midway, above right] {No};
        \draw[line] (left) -- (h2) node[midway, above left] {Yes};
        \draw[line] (left) -- (h3) node[midway, above right] {No};
    
        \end{tikzpicture}
    \end{minipage}%
    \begin{minipage}{.5\linewidth}
      \centering
      \begin{tikzpicture}[scale=4] 
            \draw[thick] (0, 0) rectangle (1, 1);
            
            \draw[thin] (0.7, 0) -- (0.7, 1) ;
            
            \draw[thin] (0, 0.6) -- (0.7, 0.6);
            
            \node[below] at (0.5, -0.1) {$x_1$};
            \node[left] at (-0.1, 0.5) {$x_2$};
    
            \node[right] at (1.05, 0.6)  {\small{$x_2 = 0.6$}};
            \node[above] at (0.7, 1.05)  {\small{$x_1 = 0.7$}};
            
            \node at (0.85, 0.5) {$h_1$};
            \node at (0.35, 0.3) {$h_2$};
            \node at (0.35, 0.8) {$h_3$};
    
            \draw[dashed] (0.85, 0.5) circle [radius=0.12];
            \draw[dashed] (0.35, 0.3) circle [radius=0.12];
            \draw[dashed] (0.35, 0.8) circle [radius=0.12];
            
        \end{tikzpicture}
    \end{minipage}
    \caption{Example of a decision tree with the corresponding partition of the covariate space $[0,1]^2$ and the associated step heights $\H = \{h_1, h_2, h_3\}$.}
    \label{ExampleTree}
\end{figure}

We place an independent prior on each tree $(\mathcal{T}_j,\mathcal{H}_j)$ in the ensemble.
The prior factorises as $p(\mathcal{T}_j,\mathcal{H}_j)=p_{\mathcal{T}}(\mathcal{T}_j)\,p_{\mathcal{H}}(\mathcal{H}_j\mid \mathcal{T}_j)$ for $j=1,\ldots,m$.
The structure prior $p_{\mathcal{T}}$ is specified as a heterogeneous Galton--Watson branching process \citep{TheoryBART}. 
Conditionally on $\mathcal{T}_j$, standard BART assigns independent Gaussian priors to the step heights, $h_{j\ell}\sim \mathcal{N}\!\left(0,1/(4\sqrt{m})\right)$. 
The scaling is chosen so that the induced prior on the sum of trees places most mass within the observed response range.
This conjugate choice enables efficient Gibbs updates. 
We modify this prior to accommodate continuous shrinkage of the step height parameters.

\subsection{Horseshoe prior on the step heights}\label{stepheight_prior}

We present our novel formulation of step height prior based on a global-local shrinkage prior.
Given a tree $\T$ with $L$ leaf nodes, we specify a prior on the set of step heights $\H = \{h_1, h_2, \dots, h_L\}$.
To achieve shrinkage in high-dimensional covariate spaces, we use a class of continuous shrinkage priors on the set of step heights. 
This class can be represented as a scale mixture of normal distributions.
In general form, the prior $p_\mathcal{H}$ is specified hierarchically:
\begin{equation}
\begin{split}
h_\ell \mid \lambda_\ell^2, \tau^2,  \omega &\sim \mathcal{N}(0, \omega \lambda_\ell^2 \tau^2), \\
\lambda_\ell^2 &\sim p(\lambda), \\
\tau^2 &\sim p(\tau),
\end{split}
\end{equation}
where $\ell = 1, \ldots, L$, and $\omega$ is a fixed hyperparameter that ensures appropriate scaling of the prior variance relative to the tree-based model structure.

The horseshoe prior \citep{OriginalHorseshoe, PolsonScott2011} arises by specifying half-Cauchy distributions for both the global and local shrinkage parameters:
\begin{equation}\label{Horseshoe}
    \tau \sim \mathcal{C}^+(0, \alpha_\tau) \quad \text{and} \quad \lambda_\ell \sim \mathcal{C}^+(0, \alpha_\lambda) \ \text{ for } \ \ell = 1, \ldots, L,
\end{equation}
where $\mathcal{C}^+(0, \alpha)$ denotes the half-Cauchy distribution with location zero and scale parameter $\alpha > 0$.
Note that the median of the distribution equals $\alpha$, while the mean and variance do not exist due to the heavy tails.
We refer to this specification of the prior as a \textsl{Horseshoe Forest}.
A Horseshoe Forest refers specifically to a single learner for the outcome, as in~\eqref{bart_model}.
In contrast, the \textsl{Causal Horseshoe Forest} decomposes the outcome into a prognostic function and a treatment effect function (see~\eqref{model}), each modelled by its own Horseshoe Forest.

This formulation falls within the class of global-local shrinkage priors \citep{PolsonScott2011}. 
The \textit{global shrinkage parameter} $\tau$ governs the overall degree of shrinkage, encouraging all step heights to be small, while the \textit{local shrinkage parameters} $\lambda_\ell$ allow individual step heights $h_\ell$ to deviate from this global tendency when supported by the data. 
In the tree-based context, the local shrinkage parameters refer to the values assigned to individual leaf nodes, each associated with a distinct region of the covariate space, whereas the global shrinkage parameter is shared across all leaf nodes. 
This hierarchical structure enables the model to suppress noise in regions lacking signal while preserving flexibility in regions where effects are present. 
The horseshoe prior is particularly well-suited for this task, as it can adapt to varying degrees of sparsity without imposing excessive shrinkage on substantial signals \citep{vdPas2017, vdPas2017Horseshoe, bhadra2019lasso}.

The global shrinkage parameter $\tau$ can be specified either as a single parameter shared across the entire forest or assigned separately per tree.
We adopt the latter approach and assign a separate global shrinkage parameter to each tree. 
This choice is motivated by two considerations. 
First, assigning independent global shrinkage parameters preserves the prior independence between trees.
While the trees are not independent in the posterior due to their shared dependence on the residuals, this prior structure encourages each tree to act as a weak learner, consistent with the additive ensemble construction underlying boosting-type methods.
Weak learners have limited individual accuracy and are combined to form a strong ensemble learner \citep{Schapire1990WeakLearnability, Freund1997Boosting}.
Second, each tree partitions the covariate space differently and may focus on distinct regions with varying signal strength. 
A shrinkage parameter shared across the forest would force the same level of regularisation across the entire covariate space.
This would be inefficient in heterogeneous settings.
Tree-specific shrinkage allows the model to adapt locally: it can apply more shrinkage in noisy regions and less in informative ones.

We prefer continuous shrinkage over exact sparsity in high-dimensional causal inference.
Model selection methods like LASSO-based methods \citep{Belloni2014, Shortreed2017} or DART \citep{linero2018dirichlet} induce sparsity by forcing some coefficients to be exactly zero, whereas shrinkage priors such as the horseshoe continuously pull estimates toward zero without fully excluding the respective covariates. 
This distinction is particularly relevant in causal settings, where omitting covariates risks violating the unconfoundedness assumption (see \eqref{unconfoundedness}).

From a conceptual perspective, the use of continuous shrinkage priors in tree-based models is also supported by their connection to the classical normal means problem \citep{james1961estimation}. 
If the covariate space were partitioned into $n$ leaf nodes, each containing a single observation, the model would reduce to a set of independent normals centred at the leaf-specific means. 
Although such fine partitions are unlikely under the prior, the Bayesian tree ensemble can be viewed as performing Bayesian model averaging over normal means models that cluster observations based on covariate similarity.
This connection also supports recent advocacy for Bayesian model averaging in causal inference \citep{Wang2012, Zigler2016, Horii2021BMA, Antonelli2020Averaging}.
This perspective becomes explicit in Section~\ref{subsection:full_conditional}, where we describe a full conditional posterior draw of the step height parameters.

\subsection{Choice of hyperparameters}\label{default_hyperparameters}
We follow the default recommendations of \citet{BARToriginal} regarding the hyperparameters governing the prior on the tree structure. 
Empirical evaluations suggest that the influence of these parameters is negligible in our setting. 
We set the number of trees to 200 for both the prognostic and treatment effect forests. 
This contrasts with the recommendation of \citet{hahn2020bayesian}, who advocate using a smaller number of trees for the treatment effect model as a form of regularisation. 
In our approach, regularisation is instead achieved through the prior on the step heights. 
We have found that using a larger number of trees, which induces a finer partition of the high-dimensional covariate space, improves estimation in this context.

We adopt an empirical approach to set the prior hyperparameters, following the recommendation of \citet{BARToriginal}. 
Specifically, we centre and standardise the log-transformed survival times using preliminary estimates of the mean and variance, obtained either from a linear AFT model for high-dimensional data \citep{Kumari2025afthd} or from an intercept-only AFT model. 
For the Horseshoe Forest model, which uses a single forest, we set the variance scaling parameter to $\omega = 1$. 
In the Causal Horseshoe Forest model, where the outcome is modelled as the sum of two forests for $f(x)$ and $\tau(x)$, we set $\omega = \frac{1}{2}$ to ensure that sufficient prior mass is allocated across the observed outcome range.

The global and local shrinkage parameters $\tau$ and $\lambda$ are assigned half-Cauchy priors with scale parameters $\alpha_\tau$ and $\alpha_\lambda$, respectively. 
For simplicity, we set $\alpha := \alpha_\tau = \alpha_\lambda$. 
Some studies recommend using a different prior scale for the global parameter $\tau$ than for the local parameters $\lambda_\ell$ to better control overall shrinkage \citep{Piironen2017Hyperprior}. 
In our experiments, we found that setting the same scale for both worked well in practice.
This fully Bayesian specification remains adaptive and robust across various settings.
We parametrise the scale as $\alpha = \frac{k}{\sqrt{m}}$ for some $k > 0$, where $m$ denotes the number of trees in the respective forest. 
The choice of $k$ governs the level of shrinkage.
We recommend setting $k = 1$ for conservative shrinkage, $k = 0.1$ for moderate shrinkage, and $k = 0.05$ for more aggressive shrinkage. 
At the start of the chain, a small value for $k$ prevents the sampler from becoming trapped in local regions. 
As the chain converges toward its stationary distribution, the heavy-tailed nature of the horseshoe prior allows the posterior to explore the full parameter space. 
In practice, we recommend selecting $k$ via cross-validation.
We perform cross-validation for censored data using the concordance index (C-statistic) \citep{harrell1982evaluating}, computed on held-out test predictions.

\subsection{Censored and binary outcomes}\label{cens_bin}

We account for censoring through data augmentation within the Gibbs sampler \citep{TannerWong1987}. 
Censored event times are treated as latent variables and sampled from their full conditional posterior distributions at each iteration. 
We adopt this strategy in both the Horseshoe Forest and Causal Horseshoe Forest to account for censoring in survival outcomes.

The model also supports binary outcomes.
We follow the probit approach of \citet{BARToriginal}, which uses a latent Gaussian variable linked to each binary response.
This latent variable is modelled as a sum of trees, using the data augmentation scheme of \citet{AlbertChib1993}.
We implement this binary extension only for the Horseshoe Forest and use it to estimate the propensity score.
\section{Posterior inference}\label{sec:sampling}

Posterior inference is performed using a Markov Chain Monte Carlo (MCMC) algorithm. 
We extend the standard Metropolis-within-Gibbs sampler used in Bayesian additive regression trees to a reversible jump-within-Gibbs scheme \citep{ReversibleJump}.
This extension is required because the global--local horseshoe prior on the step heights breaks the conjugacy used in the original BART algorithm \citep{BARToriginal}. 
The original independent Gaussian prior allows analytical integration over the step heights. 
This enables separate updates of the tree structures and step heights.
We instead use a reversible jump Metropolis--Hastings step to jointly update both the tree structure and the step heights. 
Each MCMC iteration alternates between updating the prognostic and treatment effect forests using a Bayesian backfitting strategy \citep{Backfitting}.

In the remainder of this section:

\begin{enumerate}
    \item \textbf{Outer Gibbs sampler.}  
    We describe the Gibbs sampling strategy for updating the components of the decomposed AFT model. Separate forests are specified for the prognostic and treatment effect functions, and survival times are handled using data augmentation.

    \item \textbf{Reversible jump step.}  
    We detail the reversible jump Metropolis--Hastings step for updating a single tree. Our approach introduces a novel proposal mechanism for the step heights and local shrinkage parameters. These candidate values are drawn from pseudo-Gibbs updates informed by the parent node, which improves efficiency and mixing.

    \item \textbf{Full conditional updates.}  
    We describe the full conditional posterior updates for the step height parameters, conditional on a fixed tree structure. This includes efficient matrix-based sampling and closed-form updates for the global and local shrinkage parameters under the horseshoe prior. We highlight the connection to classical normal means models, which provides intuition for this structure.
\end{enumerate}

We provide details on computational cost, scalability, and empirical runtimes in the Supplementary Materials.

\subsection{Outer Gibbs sampler}\label{sec:outer_gibbs}
For both the prognostic regression function $f(x)$ and the treatment effect regression function $\tau(x)$, we specify separate Horseshoe Forests. 
We denote the ensemble for $f(x)$ by $\big(\mathcal{T}_j^f, \mathcal{H}_j^f\big)_{j=1}^{m_f}$ and for $\tau(x)$ by $\big(\mathcal{T}^\tau_j, \mathcal{H}^\tau_j\big)_{j=1}^{{m_\tau}}$, where $m_f$ and ${m_\tau}$ represent the number of trees in each respective forest. 
The outer Gibbs sampler is summarised in Algorithm~\ref{OuterGibbs}.

The trees in each forest are iteratively updated conditional on all other trees. 
This conditional updating step depends on the current model state only through the $n$ residuals, which are derived from the model in Eq.~\eqref{model}.
Specifically, the residuals for the $j$-th tree in the prognostic forest and treatment effect forest are given by:
\begin{subequations}\label{Residuals}
\begin{align}
\mathcal{R}_j^f &:= \log T^o - \sum_{J=1,\, J\neq j}^{m_f} g(X; \mathcal{T}_J^f, \mathcal{H}_J^f) - A \sum_{J=1}^{{m_\tau}} g(X; \mathcal{T}^\tau_J, \mathcal{H}^\tau_J), \label{Residuals_f} \\
\mathcal{R}^\tau_j &:= \log T^o - \sum_{J=1}^{m_f} g(X; \mathcal{T}_J^f, \mathcal{H}_J^f) - A \sum_{J=1,\, J\neq j}^{{m_\tau}} g(X; \mathcal{T}^\tau_J, \mathcal{H}^\tau_J), \label{Residuals_tau}
\end{align}
\end{subequations}
where $T^o$ denotes the observed or imputed survival times after data augmentation. 
We refer to $\mathcal{R}_j^f$ as the \textsl{prognostic residuals} and to $\mathcal{R}_j^\tau$ as the \textsl{treatment effect residuals}.

Algorithm~\ref{OuterGibbs} produces posterior draws of the conditional average treatment effect by evaluating differences of marginal conditional expectations under $A=1$ and $A=0$ given the current model parameters.
It does not generate paired potential outcomes $(T_i(1),T_i(0))$ for individual units.
All stochastic updates in the outer Gibbs sampler depend on the observed (or augmented) factual outcomes $T^o$ exclusively through the residuals defined in \eqref{Residuals}.
The procedure targets marginal conditional means and does not assume or attempt to recover a joint distribution for individual-level potential outcomes.
Inference on individual causal contrasts would require additional modelling assumptions on the joint distribution of the potential outcomes \citep{Oganisian2025Untangling, Li2023BayesianCausal}.

The censored survival times are augmented within the Gibbs sampler. 
At each iteration, these times are imputed by drawing from their conditional distribution. 
This conditional distribution is a truncated normal, with the lower bound given by the observed censoring time. 
The framework can be naturally extended to interval-censored data, in which case the truncated normal is bounded by the corresponding interval limits. 
The truncated normal distribution is centred at the current model-based prediction $\widehat{T}$ of the survival time, computed using \eqref{model}, and has variance equal to the current draw of $\sigma^2$.
Detailed derivations and further discussion can be found in \citep{TannerWong1987}.

The error variance $\sigma^2$ is sampled from its full conditional distribution.
The inverse-$\chi^2$ prior is conjugate.
This yields an inverse-gamma update based on the residuals $r_i = \log{T^o_i} - \log{\widehat{T}_i}$. 
Detailed derivations can be found in \citet[Section 3.2 and 14.2]{Gelman2013BDA}.

We adopt the conventional binary treatment indicator $A$, coded as 1 for treated individuals and 0 for controls, to clearly communicate our model structure and causal assumptions. 
However, the proposed model can also accommodate the invariant parametrisation introduced by \citet{hahn2020bayesian}, which relies on a data-adaptive treatment coding scheme.

\begin{algorithm}[tb]
\caption{Outer Gibbs sampler -- $(t+1)$-th iteration}
\label{OuterGibbs}
\KwIn{Previous $\big(\mathcal{T}_j^f, \mathcal{H}_j^f\big)^{(t)}$ for $j=1,\ldots,m$, $\big(\mathcal{T}_j^\tau, \mathcal{H}_j^\tau\big)^{(t)}$ for $j=1,\ldots,\Tilde{m}$, $T_i^o$ for $i=1,\ldots,n$, $\sigma^2{}^{(t)}$}

\vspace{1ex}

\For{$j = 1$ \KwTo $m^f$}{
    Compute prognostic residuals $\mathcal{R}^f_j{}^{(t)}$ using Eq.~\eqref{Residuals_f}\;
    Sample $\big(\mathcal{T}_j^f, \mathcal{H}_j^f\big)^{(t+1)}$ from $\big(\mathcal{T}_j, \mathcal{H}_j\big) \mid \mathcal{R}_j^{f(t)}, \sigma^2$\;
}

\vspace{1ex}

\For{$j = 1$ \KwTo $m^\tau$}{
    Compute treatment effect residuals $\mathcal{R}_j^{\tau(t)}$ using Eq.~\eqref{Residuals_tau}\;
    Sample $\big(\mathcal{T}_j^\tau, \mathcal{H}_j^\tau\big)^{(t+1)}$ from $\big(\mathcal{T}_j, \mathcal{H}_j\big) \mid \mathcal{R}_j^{\tau(t)}, \sigma^2$\;
}

\vspace{1ex}

\For{$i = 1$ \KwTo $n$}{

    \eIf{$\delta_i = 0$}{
        Compute survival time prediction $\widehat{T}^{(t)}_i$using Eq.~\eqref{model}\;
        Sample $T^o_i{}^{(t+1)}$ from $T^o_i \mid Y_i, \widehat{T}^{(t)}_i, \sigma^2{}^{(t)}$\;
    }{
        Set $T^o_i{}^{(t+1)} = Y_i$\;
    }
}

\vspace{1ex}

Sample $\sigma^2{}^{(t+1)}$ from $\sigma^2 \mid \Big\{\big(\mathcal{T}_j^f, \mathcal{H}_j^f\big)^{(t+1)} \Big\}_{j=1}^{m^f}, \Big\{\big(\mathcal{T}_j^\tau, \mathcal{H}_j^\tau\big)^{(t+1)}\Big\}_{j=1}^{m^\tau}, \Big\{T^o_i{}^{(t+1)}\Big\}_{i=1}^n$.
\end{algorithm}

\subsection{Reversible jump step for updating a tree}\label{subsec:RJ_step}

The tree structure is jointly updated together with its corresponding set of step heights. 
We propose a new pair $(\mathcal{T}_j, \mathcal{H}_j)$, which is then accepted or rejected using a Metropolis--Hastings step with a reversible jump move \citep{ReversibleJump}.
The conditional posterior distribution of a single tree factorises as
\begin{equation}
    p\bigl((\mathcal{T}_j, \mathcal{H}_j) \mid \mathcal{R}_j, \sigma^2\bigr) \propto  p\bigl(\mathcal{H}_j \mid \mathcal{T}_j, \mathcal{R}_j, \sigma^2\bigr) \times p\bigl(\mathcal{T}_j \mid \mathcal{R}_j, \sigma^2\bigr).
\end{equation}
This factorisation enables a two-step proposal mechanism: first proposing a new tree structure $\mathcal{T}_j$, then proposing new step heights $\mathcal{H}_j$ conditional on the proposed structure. 
The overall proposal is evaluated using the reversible jump acceptance probability to ensure detailed balance.

\begin{figure}[tb]
    \centering
\begin{tikzpicture}[scale=3] 
    \draw[thick] (0, 0) rectangle (1, 1);
    
    \draw[thin] (0.7, 0) -- (0.7, 1) ;
    
    \draw[thin] (0, 0.6) -- (0.7, 0.6);
    
    \node[below] at (0.5, -0.1) {$x_1$};
    \node[left] at (-0.1, 0.5) {$x_2$};

    
    \node at (0.85, 0.5) {$h_1$};
    \node at (0.35, 0.3) {$h_2$};
    \node at (0.35, 0.8) {$h_3$};

    \draw[dashed] (0.85, 0.5) circle [radius=0.10];
    \draw[dashed] (0.35, 0.3) circle [radius=0.10];
    \draw[dashed] (0.35, 0.8) circle [radius=0.10];
    
    \begin{scope}[shift={(2.0, 1.25)}]
        \draw[thick] (0, 0.00) rectangle (1, 1);
    
        \draw[thin] (0.7, 0) -- (0.7, 1);
        \draw[thin] (0, 0.6) -- (0.7, 0.6);
        \draw[thin] (0.4, 0.6) -- (0.4, 1.0);
        
        \node at (0.85, 0.5) {$h_1$};
        \node at (0.35, 0.3) {$h_2$};
        \node at (0.55, 0.8) {$h_3'$};
        \node at (0.2, 0.8) {$h_4'$};
    
        \draw[dashed] (0.85, 0.5) circle [radius=0.10];
        \draw[dashed] (0.35, 0.3) circle [radius=0.10];
        \draw[dashed] (0.55, 0.8) circle [radius=0.10];
        \draw[dashed] (0.2, 0.8) circle [radius=0.10];
    \end{scope}
    
    \begin{scope}[shift={(2.0, 0.0)}]
        \draw[thick] (0, 0.0) rectangle (1, 1.0);
    
        \draw[thin] (0.7, 0) -- (0.7, 1);
        
        \node at (0.85, 0.5) {$h_1$};
        \node at (0.35, 0.5) {$h_2'$};
    
        \draw[dashed] (0.85, 0.5) circle [radius=0.10];
        \draw[dashed] (0.35, 0.5) circle [radius=0.10];
    \end{scope}
    
    \begin{scope}[shift={(2.0, -1.25)}]
        \draw[thick] (0, 0.0) rectangle (1, 1.0);
        
        \draw[thin] (0.7, 0) -- (0.7, 1) ;
        \draw[thin] (0.3, 0.0) -- (0.3, 1);

        \node at (0.85, 0.5) {$h_1$};
        \node at (0.15, 0.5) {$h_2'$};
        \node at (0.5, 0.5) {$h_3'$};
    
        \draw[dashed] (0.85, 0.5) circle [radius=0.10];
        \draw[dashed] (0.15, 0.5) circle [radius=0.10];
        \draw[dashed] (0.5, 0.5) circle [radius=0.10];
    \end{scope}
    
    \draw[->, thick] (1, 0.5) -- (2.0, 1.75) node[midway, sloped, above] {\texttt{GROW}}; 
    \draw[->, thick] (1, 0.5) -- (2.0, 0.5) node[midway, sloped, above] {\texttt{PRUNE}}; 
    \draw[->, thick] (1, 0.5) -- (2.0, -0.75) node[midway, sloped, above] {\texttt{CHANGE}};
\end{tikzpicture}
\caption{Examples of covariate space partitions induced by each of the three moves on the tree in Figure \ref{ExampleTree}, with corresponding step heights.}

    \label{NewTrees}
\end{figure} 

The proposal mechanism relies on three fundamental move types: \texttt{GROW}, \texttt{PRUNE}, and \texttt{CHANGE} \citep{BayesianCART}. 
Figure~\ref{NewTrees} schematically illustrates these moves and shows how the dimensionality of the step height parameter vector $\mathcal{H}_j$ changes accordingly. 
In a \texttt{GROW} move, the tree expands by adding a split, introducing additional step height parameters (and corresponding local horseshoe variance parameters). 
Conversely, a \texttt{PRUNE} move reduces the tree size by removing a split and eliminating corresponding parameters. 
A \texttt{CHANGE} move modifies an existing split without altering the number of terminal nodes.

Along with each move, new values for the step height parameters are proposed. 
The idea is to mimic a Gibbs sampler locally: propose new parameters informed by current values, as if the dimension were fixed. 
While this pseudo Gibbs proposal does not directly sample from the true posterior, it provides a reasonable approximation.
This is corrected using the Metropolis--Hastings acceptance step to leave the posterior invariant.

We illustrate the proposal mechanism for the \texttt{GROW} move.
Details for the \texttt{PRUNE} and \texttt{CHANGE} moves are provided in Supplementary Materials~\ref{appendix:comp_details} \citep{supplement}.
Suppose we perform a \texttt{GROW} move on leaf $L$.
Then two new child leaves $L$ and $L+1$ are grown.
We propose new values $h'_L$, $h'_{L+1}$, $\lambda'_L$, and $\lambda'_{L+1}$ by drawing from conditional distributions informed by the parent node parameters $h_L$ and $\lambda_L$.
These conditional distributions can be derived using Bayes' rule.
We do not need to propose a global shrinkage parameter $\tau$ since this dimensionality is not affected by the tree moves.

The horseshoe prior can be rewritten using an auxiliary variable representation \citep{HSGibbs}:
\begin{equation}\label{AuxilliaryHorseshoe}
\begin{alignedat}{2}
\tau^2 \mid \xi         &\sim \mathcal{IG}\left(\frac{1}{2}, \frac{1}{\xi}\right) &\quad \quad\text{and}\quad\quad  \lambda_\ell^2 \mid \nu_\ell &\sim \mathcal{IG}\left(\frac{1}{2}, \frac{1}{\nu_\ell}\right),\\
\xi   &\sim \mathcal{IG}\left(\frac{1}{2}, \frac{1}{\alpha^2}\right) &  \nu_\ell &\sim \mathcal{IG}\left(\frac{1}{2}, \frac{1}{\alpha^2}\right).
\end{alignedat}
\end{equation}
for $\ell = 1, \ldots, L$.
This representation allows all conditional distributions to be derived in closed form.
For notational simplicity, we write $\cdot$ in the conditioning statement to indicate dependence on all other model parameters and the data.
For the new node $L$, the new parameters are sequentially proposed as:
\begin{equation}
\begin{split}
\nu_L' \mid \cdot &\sim \mathcal{IG}\left(1, \frac{1}{\alpha} + \frac{1}{\lambda_L^2}\right),\\
\lambda_L'{}^2 \mid \cdot &\sim \mathcal{IG}\left(1, \frac{1}{\nu_L'} + \frac{h_L^2}{2\tau^2\omega}\right),\\
h_L' \mid \cdot &\sim \mathcal{N}\left( \frac{\bar{\mathcal{R}}_L'}{n_L' + \frac{1}{\tau^2\lambda_L'{}^2\omega}}, \frac{\sigma^2}{n_L' + \frac{1}{\tau^2\lambda_L'{}^2\omega}}\right),
\end{split}
\end{equation}
where $\bar{\mathcal{R}}_L'$ and $n_L'$ denote the sum and count of residuals mapped to child node $L$ by the proposed tree.  
The same procedure is applied to propose the parameters for node $L+1$.

A summary of the single-tree update is given in Algorithm~\ref{SingleTreeUpdate}.
Detailed acceptance ratios and derivations are provided in Supplementary Materials~\ref{appendix:comp_details}.
Related RJMCMC-based approaches for BART models in settings where step heights cannot be analytically integrated out have recently been studied by \citet{GeneralizedBART}.
While our modelling context and shrinkage structure differ, both approaches rely on joint proposals of tree structures and associated step heights within an RJMCMC framework.
\begin{algorithm}[tb]
\caption{Update of a single tree via reversible jump step}
\label{SingleTreeUpdate}

\KwIn{$(\mathcal{T}_j, \mathcal{H}_j)$, $\mathcal{R}_j$, $T_i^o$ for $i=1,\ldots,n$, $\sigma^2$}

\vspace{1ex}

Randomly select and perform one of the moves: \texttt{GROW}, \texttt{PRUNE}, or \texttt{CHANGE}\;

\vspace{1ex}

Propose new step height parameters corresponding to the proposed tree structure\;

\vspace{1ex}

Compute the reversible jump acceptance ratio\;

\vspace{1ex}

Draw $U \sim \mathcal{U}[0,1]$ and accept or reject the proposed $(\mathcal{T}_j, \mathcal{H}_j)$ accordingly\;

\vspace{1ex}

Perform a full Gibbs update of the step height parameters $\mathcal{H}_j$ as described in Section~\ref{sec:sampling}.

\end{algorithm}

\subsection{Full conditional update of step height parameters}\label{subsection:full_conditional}

After performing a reversible jump Metropolis--Hastings step that acts locally on a specific terminal node of the tree $\mathcal{T}_j$, we carry out a full conditional update of the associated step height parameters $\mathcal{H}_j$. 
Conditional on the current tree structure, the dimensionality of the parameter space remains fixed, enabling efficient Gibbs sampling updates. 
Given a tree $\mathcal{T}$ and observed covariates $x_1, \ldots, x_n$, the $n$ observations are partitioned into $L$ step heights. 
This partition can be encoded by a design matrix $\mathcal{D} \in \mathbb{R}^{n \times L}$, where each row contains exactly one entry of 1 and zeros elsewhere, indicating the assigned step height. 
Specifically, if observation $i$ belongs to leaf $\ell$, then $\mathcal{D}_{i\ell} = 1$. 
For $\mathcal{D}=I_n$ this reduces to a normal means model.
By the valid partitions assumption, each row of $\mathcal{D}$ contains at least one non-zero entry, ensuring no empty leaves.

This structure allows us to reformulate the update as a regression problem in terms of the step heights $\smash{h = (h_1, \ldots, h_L)}$:
\begin{equation}
    \mathcal{R} \mid \mathcal{D}, h, \sigma^2 \sim \mathcal{N}\left(\mathcal{D}h, \sigma^2 I_n\right),
\end{equation}
where $\mathcal{R} \in \mathbb{R}^n$ denotes the residuals as defined in (\ref{Residuals}).

The conditional posterior of $h \in \mathbb{R}^L$ is given by \citep{lindley1972bayes}:
\begin{equation}
    h \sim \mathcal{N}_L\left(\Omega^{-1}\mathcal{D}^\top \mathcal{R}, \Omega^{-1}\right), \quad \text{where} \quad \Omega = \mathcal{D}^\top \mathcal{D} + \left[\omega\tau^2 \operatorname{diag}(\lambda_1^2, \ldots, \lambda_L^2)\right]^{-1}.
\end{equation}
Since $\mathcal{D}^\top \mathcal{D}$ is a diagonal matrix with leaf sizes on its diagonal, let $n_\ell$ denote the number of observations in leaf $\ell$ and $\bar{\mathcal{R}}_\ell$ the sum of residuals in leaf $\ell$. 
We can then express the conditional posterior of each $h_\ell$ as:
\begin{equation}
    h_\ell \mid \cdot \sim \mathcal{N}\left(\frac{\bar{\mathcal{R}}_\ell}{n_\ell + \frac{1}{\tau^2\lambda_\ell^2\omega}}, \frac{\sigma^2}{n_\ell + \frac{1}{\tau^2\lambda_\ell^2\omega}}\right).
\end{equation}

The conditional posteriors for the shrinkage parameters follow directly from the auxiliary parametrisation of the horseshoe prior, as presented in (\ref{AuxilliaryHorseshoe}):
\begin{equation}\label{PosteriorAuxiliary}
\begin{alignedat}{2}
\tau^2 \mid \cdot &\sim \mathcal{IG}\left(\frac{L+1}{2}, \frac{1}{\xi} + \frac{1}{2\omega} \sum_{\ell=1}^{L} \frac{h_\ell^2}{\lambda_\ell^2}\right)
&\quad\text{and}\quad&
\lambda_\ell^2 \mid \cdot \sim \mathcal{IG}\left(1, \frac{1}{\nu_\ell} + \frac{h_\ell^2}{2\tau^2\omega}\right), \\
\xi \mid \cdot &\sim \mathcal{IG}\left(1, \frac{1}{\alpha^2} + \frac{1}{\tau^2}\right)
&\quad\text{   }\quad&
\nu_\ell \mid \cdot \sim \mathcal{IG}\left(1, \frac{1}{\alpha^2} + \frac{1}{\lambda_\ell^2}\right).
\end{alignedat}
\end{equation}
All of these conditional distributions are inverse gamma, and together with the Gaussian update for $h$, enable efficient Gibbs sampling within each tree. 
This hierarchical structure facilitates scalable and adaptive updates while maintaining strong shrinkage properties, thus improving mixing and posterior exploration.

\section{Simulation studies}\label{section:Simulations}

We conduct two sets of simulation studies to evaluate the empirical performance of the proposed Causal Horseshoe Forest. 
The first set assesses general accuracy and uncertainty quantification for estimating heterogeneous treatment effects under varying levels of dimensionality.
The second set focuses on a more targeted scenario designed to study regularisation-induced confounding.

\subsection{General performance}\label{sec:sim-setup}

We compare the Causal Horseshoe Forest to several AFT-based Bayesian tree ensemble methods commonly used for heterogeneous treatment effect estimation.
We consider the following models:
{\setlength{\itemsep}{1pt}
 \setlength{\parsep}{0pt}
\begin{itemize}
    \item \textbf{Causal Horseshoe Forest}: our proposed method, with shrinkage parameter $k$ selected via 10-fold cross-validation;
    \item \textbf{AFT-BART}: an AFT extension of Bayesian additive regression trees proposed by \citet{henderson2020individualized}, which fits a single BART model to estimate the conditional average treatment effect;
    \item \textbf{AFT-DART}: an extension of AFT-BART that places a Dirichlet prior on the covariate selection probabilities \citep{linero2018dirichlet};
    \item \textbf{AFT-BCF}: an AFT extension of Bayesian Causal Forest (BCF) \citep{hahn2020bayesian};
    \item \textbf{AFT-Shrinkage BCF}: an AFT extension of Shrinkage BCF \citep{caron2022shrinkage} that assigns Dirichlet priors to the covariate selection probabilities of both tree ensembles.
\end{itemize}
} 
We implement all AFT-based benchmark methods in their semiparametric form. 
This is correctly specified for the data-generating mechanisms considered. 
We sample 5000 posterior draws after a burn-in of 2500 iterations and use 1000 Monte Carlo replications across all scenarios.
We evaluate each method’s ability to estimate the conditional average treatment effect (CATE) using the root mean squared error (RMSE), empirical coverage probability, and average credible interval length. 
The RMSE is computed across individuals as the square root of the mean squared error between the estimated and true CATE values.
The mean squared error of the CATE is also referred to as the precision in estimating heterogeneous effects (PEHE) as defined by \citet{Hill2011}.

We generate data for each individual $i = 1, \ldots, n$ using the following hierarchical structure:
\begin{equation}
    \begin{split}
        X_i &\sim \mathcal{U}[0,1]^p, \\
        A_i &\sim \text{Ber}(e(x_i)), \\
        \log T_i &\sim \mathcal{N}(f(x_i) + a_i \tau(x_i), \sigma^2), \\
        C_i &\sim \text{Exp}(\eta), \\
        Y_i &= \min(T_i, C_i),
    \end{split}
\end{equation}
where $\eta>0$ is chosen via a Monte Carlo procedure to achieve an average censoring rate of approximately 35\% and set $\sigma^2 = 3$.
We define the propensity score and the prognostic function:
\begin{equation}
\begin{split}
    e(x_i) &= \Phi(-0.5 + 0.4x_{i1} - 0.1x_{i3} + 0.3x_{i5}), \\
    f(x_i) &= \beta_f^\top x_i,
\end{split}
\end{equation}
where the entries of $\beta_f$ follow a spike-and-slab distribution given by $\beta_f \sim (1-s)\delta_{0_p} + s\mathcal{N}(0_p,I_p)$, with sparsity $s = 0.1$.
We consider two treatment effect functions:
\begin{equation}
\begin{split}
    \text{(Linear)} \quad & \tau(x_i) = 1 + x_{i1} - 2x_{i2} + 3x_{i3} - 4x_{i4} + 5x_{i5} + \beta_\tau^\top x_i, \\
    \text{(non-linear)} \quad & \tau(x_i) = 10 \sin(\pi x_{i1} x_{i2}) + 20 (x_{i3} - 0.5)^2 + 10x_{i4} + 5x_{i5},
\end{split}
\end{equation}
where the entries of $\beta_\tau$ also follow a spike-and-slab distribution with $s = 0.05$.
The linear scenario explicitly introduces confounding while keeping sparsity constant through the $\beta_\tau^\top x$ term. 
The non-linear scenario follows the benchmark Friedman function \citep{Friedman}.
In this scenario the number of contributing covariates is fixed.
As $p$ increases, the model effectively becomes sparser. 
We consider three covariate dimensions $p = {100, 1000, 5000}$ and keep the sample size fixed at $n = 200$.

We examine the non-linear scenario in greater depth in an extended simulation.
We increase the covariate dimension incrementally from $p = 50$ to $p = 5000$.
The non-linear treatment effect function is augmented with additional sparse linear main effects and sparse linear interaction terms on top of the baseline non-linear component:
\begin{equation}
\begin{split}
\tau(x_i) &=
10 \sin(\pi x_{i1} x_{i2})
+ 20 (x_{i3} - 0.5)^2
+ 10 x_{i4}
+ 5 x_{i5} + \beta_{\tau}^\top x_i
+ x_i^\top \Gamma x_i ,
\end{split}
\end{equation}
where $\Gamma \in \mathbb{R}^{p\times p}$ is a sparse symmetric interaction matrix with
$\Gamma_{jj}=0$ and $(x_i^\top \Gamma x_i)=\sum_{1\le j<k\le p}\Gamma_{jk}x_{ij}x_{ik}$.
The main-effect coefficients follow a spike-and-slab prior with sparsity $s_{\mathrm{main}}=0.05$.
The interaction coefficients are generated analogously by drawing the upper-triangular
entries $\Gamma_{jk}$ independently from a spike-and-slab distribution with sparsity
$s_{\mathrm{int}}=0.01$.

\begin{figure}[tb]
\centering
\includegraphics[width=\linewidth]{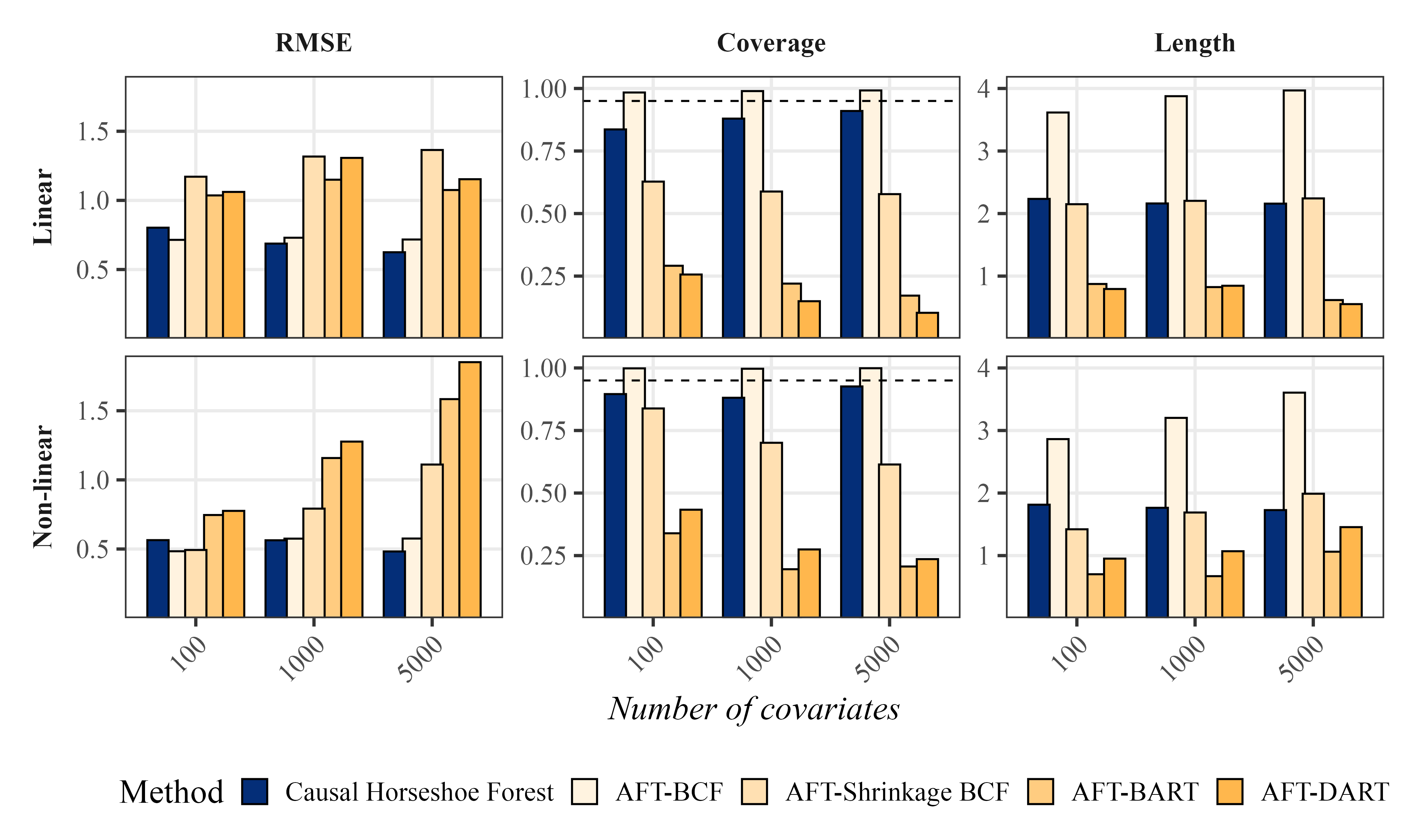}
    \caption{Simulation results for linear and non-linear treatment effect scenarios. Columns report CATE performance metrics: RMSE (left), empirical coverage (middle; dashed line indicates the nominal 95\% level), and average credible interval length (right). Results are shown for increasing covariate dimension $p = {100, 1000, 5000}$ across all methods.}
\label{fig:main_verticalbars}
\end{figure}

\subsubsection{Results}\label{sec:sim_results}

The results are presented in Figure~\ref{fig:main_verticalbars} and summarised in tabular form in Section~S.3 of the Supplementary Material. 
The Causal Horseshoe Forest and AFT-BCF attain the lowest RMSE. 
In the low-dimensional setting ($p = 100$), AFT-BCF is slightly more accurate. 
In high-dimensional settings ($p > n$), the Causal Horseshoe Forest has a small edge. 
The differences between these two methods remain modest across scenarios. 
Both methods outperform AFT-Shrinkage BCF across all considered values of $p$. 
The single-forest approaches (AFT-BART and AFT-DART) show a substantial increase in RMSE as $p$ increases.

In terms of uncertainty quantification, the Causal Horseshoe Forest achieves coverage that remains stable at approximately 90\% across $p$. 
Its credible intervals remain moderate in length and nearly constant across dimensions.
AFT-BCF achieves coverage close to one, but this is accompanied by substantially wider credible intervals. 
AFT-BART, AFT-DART, and AFT-Shrinkage BCF exhibit clear under-coverage.
The Causal Horseshoe Forest thus combines stable coverage with comparatively shorter credible intervals.

\begin{figure}[tb]
    \centering
    \includegraphics[width=\linewidth]{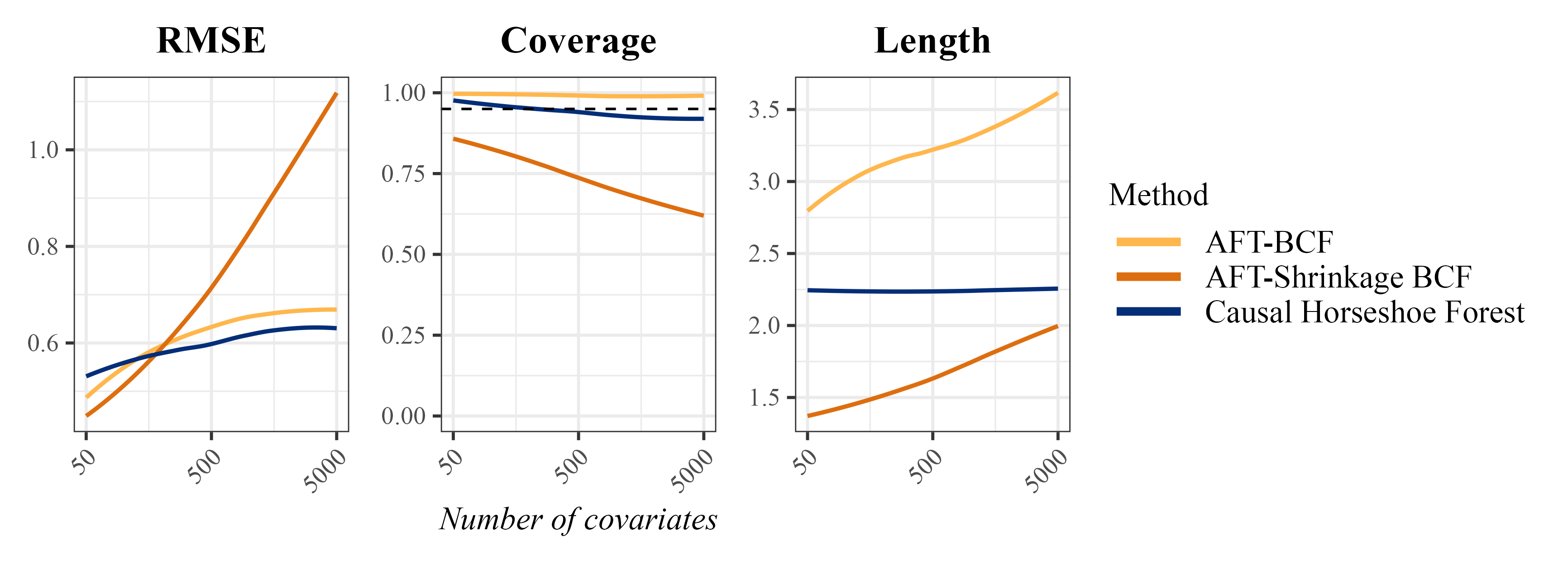}
    \caption{Performance comparison of the Causal Horseshoe Forest, AFT-BCF and AFT-Shrinkage BCF on the extended non-linear simulation. We display, from left to right, the RMSE, empirical coverage, and average credible interval length of the CATE as the covariate dimension $p$ increases from 50 to 5000.}
    \label{fig:deeper_friedman}
\end{figure}

Figure~\ref{fig:deeper_friedman} shows the comparison of the Causal Horseshoe Forest, AFT-BCF, and AFT-Shrinkage BCF based on the extended simulation study under increasing covariate dimension. 
We omit AFT-BART and AFT-DART from these plots because their RMSE increases rapidly and their coverage deteriorates substantially. 
AFT-BCF is the closest competitor to the Causal Horseshoe Forest in terms of RMSE. 
Both methods achieve similar accuracy in low-dimensional settings. 
The Causal Horseshoe Forest remains stable as the number of covariates increases. 
The main differences arise in uncertainty quantification. 
AFT-BCF achieves coverage close to one across all dimensions. 
This behaviour is driven by increasingly wide credible intervals whose length grows steadily with dimensionality. 
The Causal Horseshoe Forest attains slightly lower coverage near the nominal level and maintains substantially shorter and nearly constant interval lengths.

We provide additional simulation results in Supplementary Materials~\ref{appendix:additional_sim} to compare regularisation strategies for Bayesian tree ensembles in high-dimensional regression settings \citep{supplement}.
We consider a simple continuous outcome model without treatment or censoring, with the aim of isolating the effect of different shrinkage formulations. 
The results illustrate how the competing regularisation approaches behave as the covariate dimension increases.

\subsection{Regularisation-induced confounding}\label{sec:sim_ric}

We study regularisation-induced confounding (RIC) in a targeted simulation designed to isolate the effects of shrinkage in the presence of weak confounding. 
We consider a setting designed to isolate weak confounding, in which some covariates primarily predict treatment assignment but have only a weak association with the outcome.
These weak confounders pose a challenge for regularised estimators, since aggressive shrinkage can attenuate their contribution and induce bias even under correct model specification. 
Against this background, we examine how different regularisation strategies affect heterogeneous treatment effect estimation under weak confounding in high-dimensional settings.

We compare the Causal Horseshoe Forest to AFT-BCF and AFT-Shrinkage BCF.
These methods employ distinct regularisation mechanisms on the treatment effect function. 
We generate data for each individual $i = 1, \ldots, n$ using the same hierarchical structure as in the previous simulations, but we modify the propensity score, prognostic function, and treatment effect. 
We partition the covariates into three disjoint sets: strong confounders $\mathcal{S}$, weak confounders $\mathcal{W}$, and noise variables. 
We fix the number of strong confounders at $|\mathcal{S}| = 5$ and vary the number of weak confounders $|\mathcal{W}| = 1, 2, ..., 20$. 
We draw $X_i \sim \mathcal{U}[-1/2,1/2]^p$, fix the covariate dimension at $p = 500$, the
sample size at $n = 100$, and set $\sigma^2 = 1$.

We define the propensity score, prognostic function, and treatment effect as:
\begin{equation}
\begin{split}
    e(x_i) &= \Phi\!\left( \frac{1}{\sqrt{5 + |\mathcal{W}|}} \sum_{j \in \mathcal{S}} x_{ij}
    + \frac{2}{\sqrt{5 + |\mathcal{W}|}} \sum_{j \in \mathcal{W}} x_{ij} \right), \\
    f(x_i) &= 4\sum_{j \in \mathcal{S}} x_{ij}
    +  \sum_{j \in \mathcal{W}} x_{ij}, \\
    \tau(x_i) &= 1 + \sum_{j \in \mathcal{S}} \beta_j x_{ij},
\end{split}
\end{equation}
where the coefficients $\beta_j$ are drawn independently from a standard normal distribution. 
Treatment effect heterogeneity depends only on the strong confounders and does not involve the weak confounders.
The weak confounders are predictive of treatment assignment but have relatively small effects on the outcome, making their contribution particularly susceptible to attenuation under regularisation.

\subsubsection{Results}\label{sec:ric_results}

\begin{figure}[b!]
    \centering
    \includegraphics[width=\linewidth]{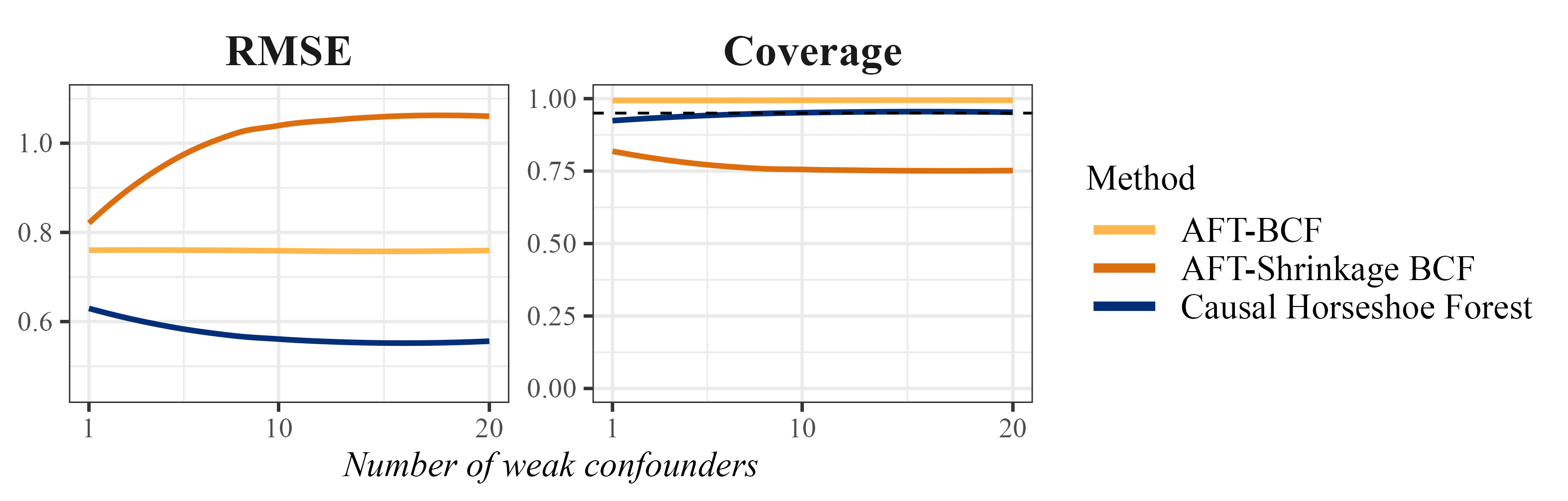}
    \caption{Regularisation-induced confounding simulation results. The figure shows RMSE (left) and empirical coverage (right) for heterogeneous treatment effect estimation as a function of the number of weak confounders. The dashed line indicates the nominal 95\% coverage level.}
      \captionsetup{font={color=Black}}
    \label{fig:ric}
\end{figure}

Continuous shrinkage on step heights provides the most stable form of regularisation for heterogeneous treatment effect estimation under different levels of weak confounding. 
Figure~\ref{fig:ric} reports RMSE and empirical coverage as functions of the number of weak confounders for the three methods considered. 
The Causal Horseshoe Forest maintains stable RMSE and near-nominal coverage as the number of weak confounders increases.
This behaviour reflects the ability of continuous shrinkage to preserve small but relevant signals and to provide well-calibrated uncertainty in high-dimensional settings.
AFT-BCF also exhibits largely constant RMSE as weak confounders are added, but coverage remains close to one across all settings. 
This suggests that regularisation through adjustment of the tree-structure priors is too weak in high-dimensional settings and results in overly wide posterior intervals.
AFT-Shrinkage BCF shows a sharp and monotone increase in RMSE as weak confounders are added, accompanied by a pronounced decline in coverage. 
The results suggest that aggressive sparsity induced by Dirichlet priors on covariate selection probabilities may exclude weak confounders, which is associated with higher bias and lower coverage under increasing regularisation-induced confounding.
\color{black}
\section{Case study: pancreatic cancer}\label{section:PDAC}

Pancreatic ductal adenocarcinoma (PDAC) is among the most aggressive and deadly cancers, characterised by rapid progression, late-stage detection, and  likely liver metastasis \citep{Pereira2019Metastasis}. 
The long-term outlook for patients remains poor, with median survival below six months and a five-year survival rate under 5\% \citep{kamisawa2016pancreatic, li2019angiogenesis}. 
PDAC exhibits substantial heterogeneity across patients, as highlighted by The Cancer Genome Atlas (TCGA) project \citep{Sidaway2017TCGA}.
Unlike many other malignancies, PDAC typically remains asymptomatic until it reaches an advanced stage. 
Surgical resection is therefore often the only curative option, typically followed by adjuvant radiation therapy. 

We study the causal effect of adjuvant radiotherapy on survival while adjusting for genetic and clinical factors that may confound or modify this effect.
We analyse data from the TCGA-PAAD cohort \citep{Sidaway2017TCGA}.
The dataset contains gene expression and clinical information for 181 PDAC tumours.
Data are curated via the \texttt{pdacR} package \citep{TorreHealy2023pdacR}.
We excluded 51 patients with missing values in either survival time, censoring status, or treatment assignment.
This leaves $ 130$ patients.
The outcome is overall survival, subject to right-censoring. 
The observed censoring rate  is 47\%.

The analysis includes several clinical covariates (see Supplementary Materials~\ref{appendix:clinical_covariates}, \citealp{supplement}) and gene expression profiles. 
A number of well-established genetic alterations drive PDAC tumorigenesis \citep{Stefanoudakis2024}.
We a priori include a subset of genes previously linked to PDAC in the literature (see Supplementary Materials~\ref{appendix:clinical_covariates}).
Additionally, we incorporate the 3{,}000 genes with the highest median absolute deviation (MAD) across patients to capture the most variable and remove low-variability genes.
This approach emphasises genes with substantial variation, which are more likely to carry informative signal for downstream modelling and reduces noise from uninformative or uniformly expressed genes \citep{Hackstadt2009, Zhang2018}.

\begin{figure}[tb]
    \centering
    \includegraphics[width=0.8\linewidth]{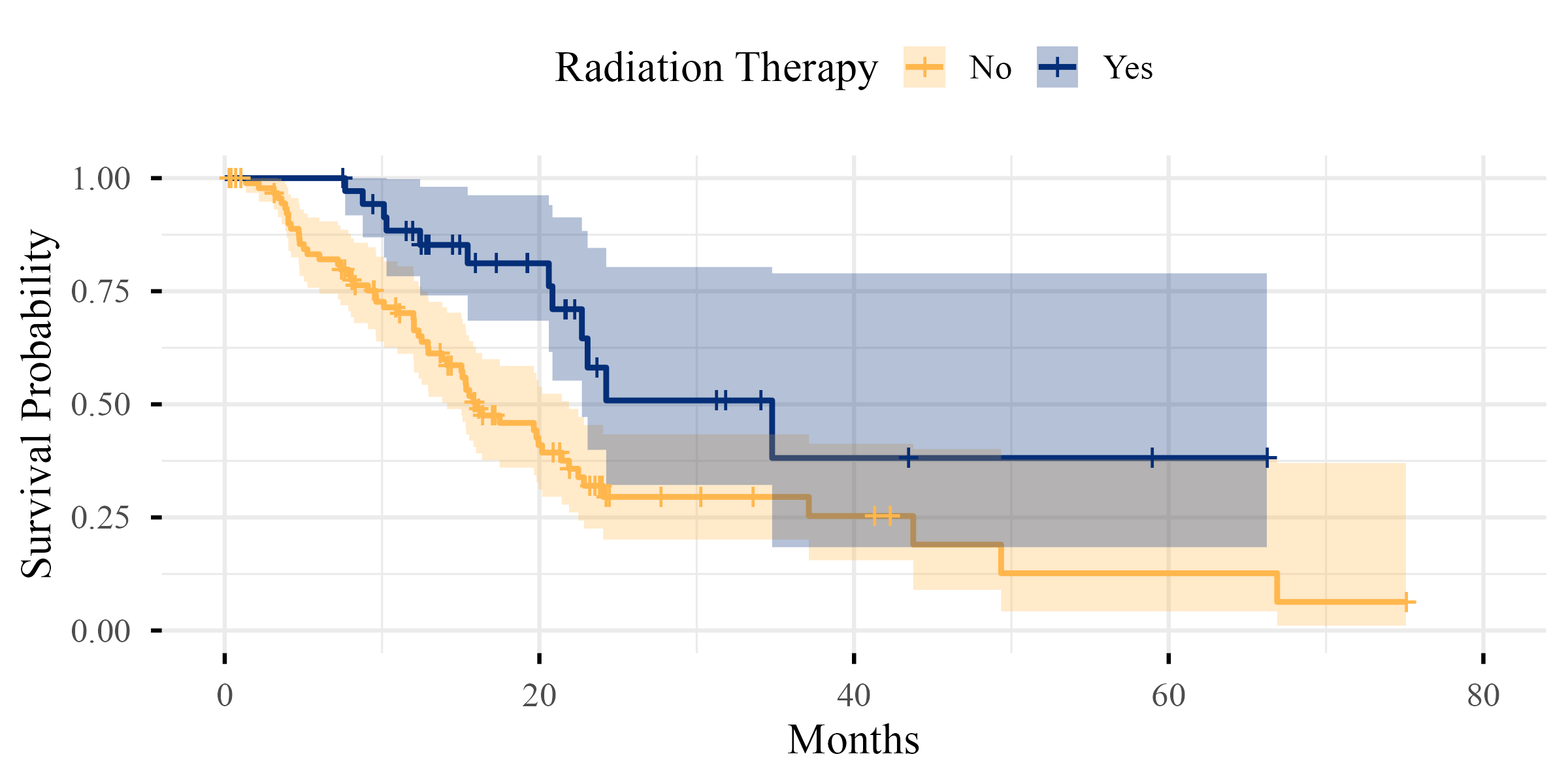}
    \caption{Kaplan--Meier survival curves for each treatment arm in the PDAC dataset.}
    \label{fig:PDAC_KaplanMeier}
\end{figure}

Median follow-up time is 23.3 months.
This is computed using the reverse Kaplan-Meier method \citep{Schemper1996}.
The estimated 2-year overall survival probability is 39.2\%.  
Figure~\ref{fig:PDAC_KaplanMeier} shows Kaplan-Meier survival curves stratified by treatment arm.

We fit a Causal Horseshoe Forest to the data with aggressive default settings in Section~\ref{default_hyperparameters}.
We performed stratified $10 \times 5$-fold cross-validation to validate the choice of hyperparameter $k$. 
Cross-validation showed no difference in terms of predictive performance across different choices of $k$ (see Supplementary Materials~\ref{appendix:clinical_covariates}).
We assessed convergence using trace plots of $\sigma^2$ (see Supplementary Materials~\ref{appendix:clinical_covariates}).
After a burn-in of 5{,}000 iterations, we retained 5{,}000 posterior samples for inference. 
The fitted model achieved a concordance index of 0.80, indicating good performance.

\begin{figure}[tb]
    \centering
        \begin{minipage}[t]{0.48\linewidth}
        \centering
        \includegraphics[width=\linewidth]{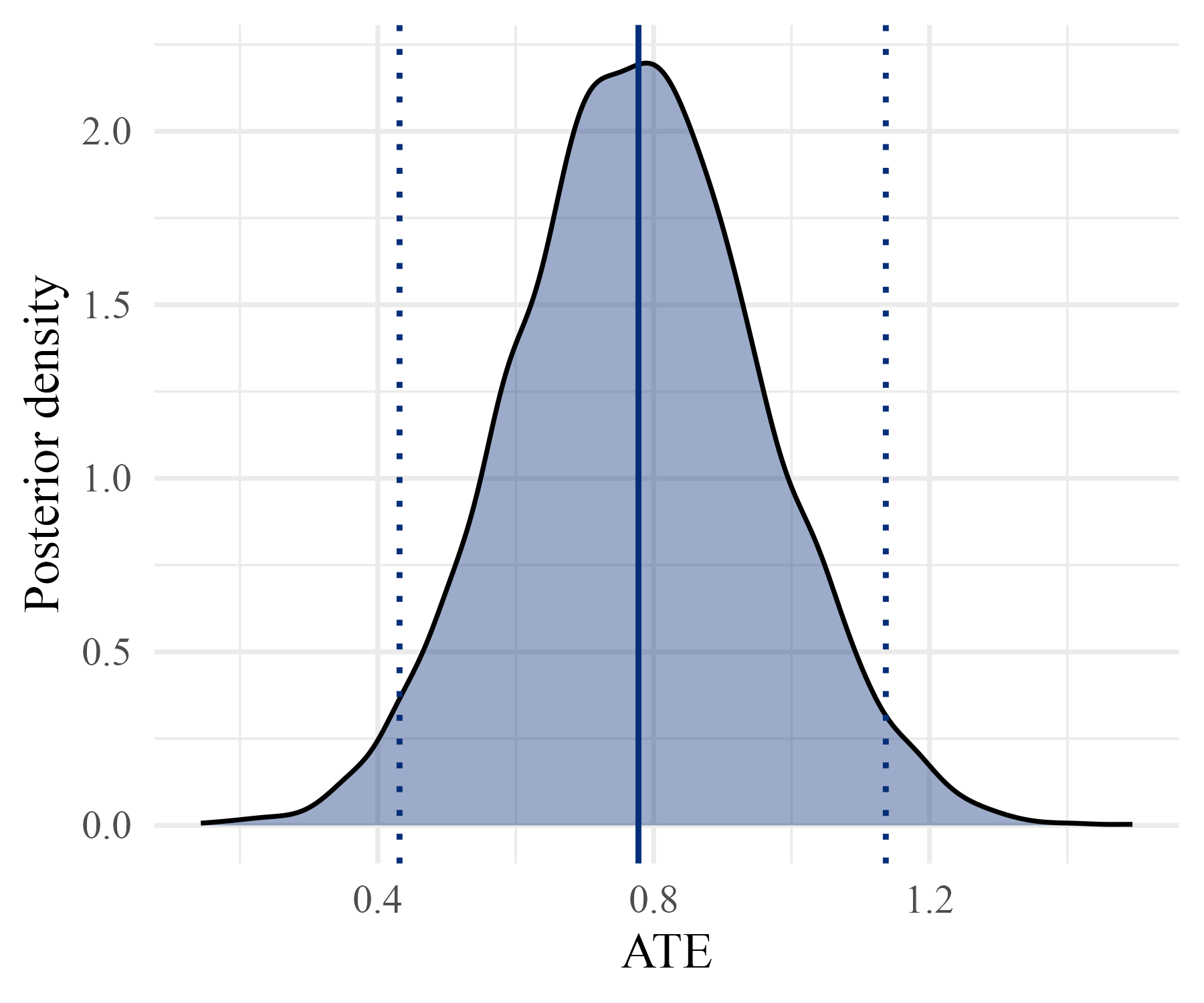}
        \caption{Posterior distribution of the ATE of adjuvant radiotherapy on overall survival in PDAC.}
        \label{fig:posterior_ATE}
    \end{minipage}
    \hfill
    \begin{minipage}[t]{0.48\linewidth}
        \centering
        \includegraphics[width=\linewidth]{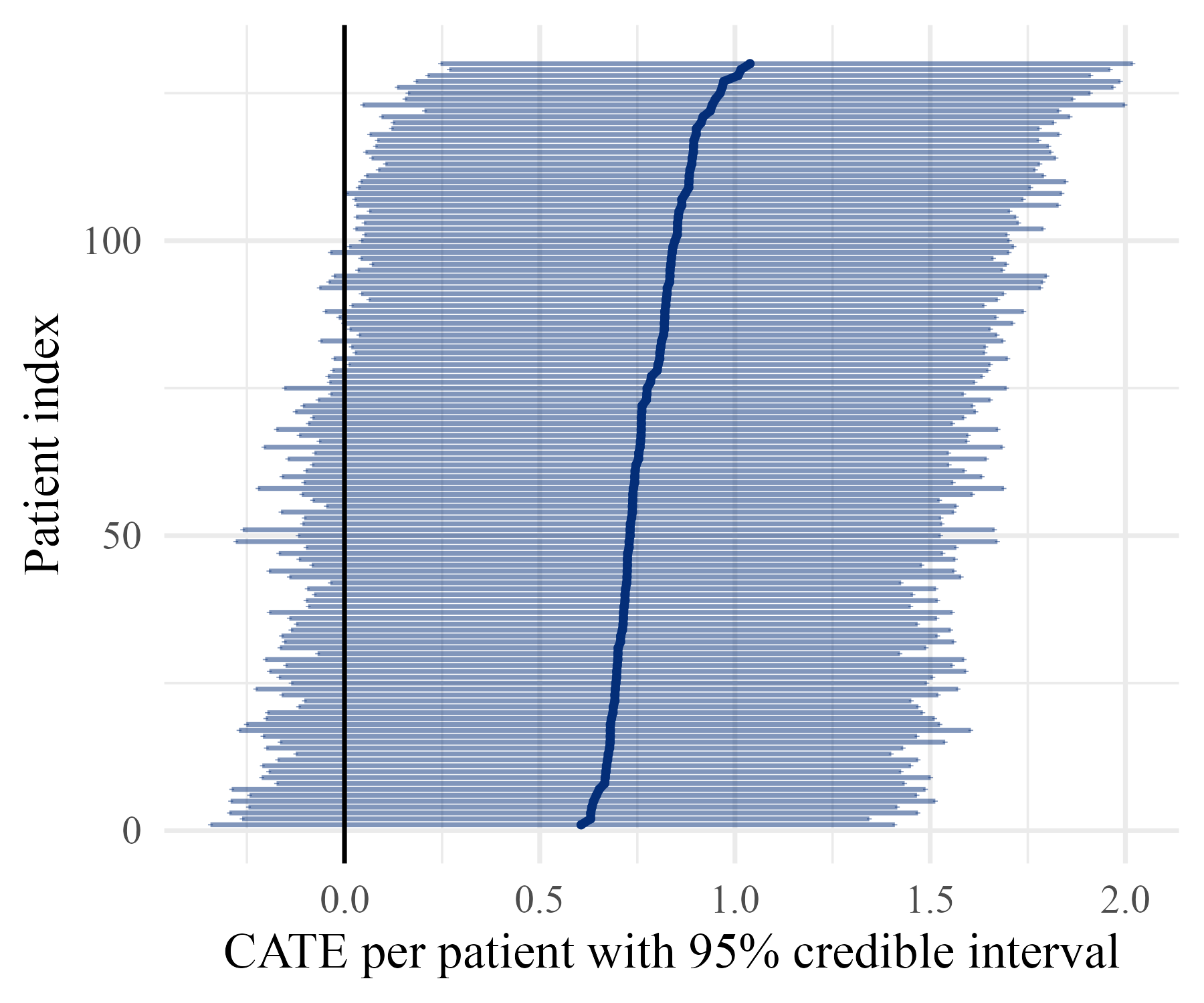}
        \caption{Posterior means and 95\% credible intervals for the individual CATEs under adjuvant radiotherapy in patients with PDAC.}
        \label{fig:cate_plot}
    \end{minipage}
\end{figure}

Figure~\ref{fig:posterior_ATE} shows the posterior distribution of the  average treatment effect (ATE, see Supplementary Materials~\ref{appendix:clinical_covariates}), with the 95\% credible interval marked by vertical dashed lines.
The posterior mean ATE is approximately 0.75, with a 95\% credible interval ranging from 0.40 to 1.11. 
This distribution is concentrated above zero, providing strong evidence for a positive average treatment effect.
These results suggest that adjuvant radiation therapy improves overall survival on average in PDAC patients.

Figure~\ref{fig:cate_plot} shows the posterior CATEs for each individual.
Each horizontal line represents a 95\% credible interval for the conditional average treatment effect.
Most intervals are wide and include zero.
This reflects substantial uncertainty and limited evidence for strong individualised effects.
A small subset of patients has intervals entirely above zero.
This suggests evidence---if any---for a positive treatment effect.
The wide intervals for many patients likely reflect insufficient information to confidently distinguish their individualised treatment effects for individuals with similar characteristics.
Nevertheless, the consistent rightward shift of the posterior means supports an overall beneficial and homogeneous effect of the treatment.

We summarise treatment effect heterogeneity using posterior probabilities of differential treatment effect and benefit \citep{henderson2020individualized}. 
For each individual, we compute 
$D_i = \p(\tau(X_i) \ge \bar\tau \mid \mathcal D)$, 
where $\bar\tau$ denotes the model-based average treatment effect and $\mathcal{D}$ the observed data. 
We transform this to 
$D_i^\ast = \max\{1 - 2D_i, 2D_i - 1\} \in [0,1]$, 
which measures posterior evidence for deviation from the population average. 
Values near $0$ indicate little evidence of heterogeneity, whereas values near $1$ indicate strong evidence. 
Population-level heterogeneity is summarised by reporting the proportion of individuals for whom $D_i^\ast$ exceeds pre-specified thresholds. 
We additionally report posterior probabilities of treatment benefit, 
$\p(\tau(X_i) > 0 \mid \mathcal D)$.

Table~\ref{tab:pdac_heterogeneity_summary} shows little evidence of substantial treatment effect heterogeneity in this cohort. 
No individuals exhibit mild or strong posterior evidence of a differential effect relative to the average treatment effect. 
At the same time, posterior probabilities of benefit are uniformly high: approximately $96.9\%$ of patients satisfy $\p(\tau(X_i) > 0 \mid \mathcal D) > 0.95$, with the remainder falling in the $(0.75, 0.95]$ range. 
These findings are consistent with earlier results: a positive average treatment effect with limited evidence for meaningful between-patient variation beyond posterior uncertainty.

\begin{table}[b!]
\centering
\begin{tabular}{lc}
\toprule
Summary measure & Percentage (\%) \\
\midrule
$D_i^\ast > 0.95$ & 0.0 \\
$D_i^\ast > 0.80$ & 0.0 \\
$\p(\tau(x_i) > 0 \mid \mathcal D) \in (0.95, 1]$ & 96.9 \\
$\p(\tau(x_i) > 0 \mid \mathcal D) \in (0.75, 0.95]$ & 3.1 \\
$\p(\tau(x_i) > 0 \mid \mathcal D) \in (0.25, 0.75]$ & 0.0 \\
$\p(\tau(x_i) > 0 \mid \mathcal D) \in [0, 0.25]$ & 0.0 \\
\bottomrule
\end{tabular}
    \caption[Posterior summaries of treatment effect heterogeneity] Posterior summaries of treatment effect heterogeneity and evidence of benefit in the PDAC analysis. The first two rows report the proportion of patients with posterior evidence of differential treatment effect relative to the average, measured by $D_i^\ast$. \citet{henderson2020individualized} suggest $D_i^\ast>0.95$ as strong evidence of differential treatment effect and $D_i^\ast>0.80$ as mild evidence.
  \label{tab:pdac_heterogeneity_summary}
\end{table}

Stable Unit Treatment Value Assumption (SUTVA) assumption is not fully met, as adjuvant radiation therapy is not standardised across patients.
Treatment regimens may vary in composition, dosage, and duration.
As a result, our analysis estimates the effect of receiving any form of radiation therapy, rather than a single, uniform intervention.
We assume sufficient overlap in a lower-dimensional confounding subspace. 
Strict overlap is unlikely in the full covariate space given the data dimensions \citep{DAmour2021Overlap}.
Positivity is fundamentally unverifiable in the $p \gg n$ regime considered here. 
Any estimated propensity score is itself a heavily regularised, high-dimensional object. 
Standard diagnostics such as propensity score histograms reflect this shrinkage behaviour as much as structural overlap. 
We therefore treat overlap as a working assumption that the data cannot directly verify.

The high mortality rate of pancreatic cancer introduces bias.
Some patients may die before becoming eligible for or completing adjuvant radiation therapy.
Thus, patients in the treatment group must have survived long enough to initiate and complete therapy.
This creates immortal time \citep{Levesque2010ImmortalTimeBias}, a period during which the event cannot occur by definition, potentially biasing results in favour of the treatment group.
We provide a preliminary landmark analysis \citep{Anderson1983TumorResponse, Putter2013Landmarking} in Supplementary Materials~\ref{appendix:landmark}.
A more thorough analysis would require detailed data on treatment timing, which are not available in this study. 
Future studies with detailed treatment timing data could help to address this bias.

\section{Discussion}\label{section:Discussion}

This work introduces a novel regularisation strategy for Bayesian regression tree ensembles by applying shrinkage directly to the step heights via the horseshoe prior. 
We use continuous global-local shrinkage to adaptively control model complexity in high-dimensional settings, while retaining all covariates in the model. 
This provides an alternative to methods that exclude covariates from the model and risk omitting important confounders.
Our method captures complex non-linear effects and higher-order interactions in both the prognostic and treatment effect functions. 
These are essential for modelling heterogeneous treatment responses in survival data.

We assume independent censoring given treatment and observed covariates. 
This assumption enables coherent likelihood-based inference via data augmentation. 
While standard in accelerated failure time models, independent censoring may be violated in practice, particularly in observational survival studies where censoring may depend on unmeasured prognostic factors. 
Inverse probability of censoring weighting \citep{RobinsFinkelstein2000} offers an alternative in such settings, but may introduce additional variance and instability. 
These issues are especially pronounced in high-dimensional settings with limited overlap and possibly extreme weights. 
Extensions of the proposed framework to explicitly accommodate informative censoring or to incorporate formal sensitivity analyses may be considered in future research.

The positivity assumption warrants further discussion.
In high-dimensional observational settings such as the PDAC application, where $p$ greatly exceeds $n$, strict positivity in the full covariate space is generically incompatible with the data \citep{DAmour2021Overlap}. 
Identification relies on overlap holding in a lower-dimensional confounding subspace, an assumption that cannot be empirically verified. 
Common diagnostics, including propensity score histograms or summaries of estimated propensity scores, are of limited use in this regime: the estimated propensity score is itself a heavily regularised, high-dimensional object.
Its distribution reflects the shrinkage behaviour of the underlying model as much as it reflects structural overlap. 
This is a fundamental limitation of high-dimensional observational causal inference rather than a feature of any particular estimator. 
We treat positivity as a working assumption throughout and emphasise that inferences from our model, like those from any causal estimator in this regime, should be interpreted with this caveat in mind.

We adopt an inclusive adjustment strategy that retains the full set of high-dimensional pretreatment covariates. 
This strategy guards against omitted variable bias, widely regarded as a primary threat to validity in observational studies \citep{Rubin2009}. 
However, conditioning on many variables may introduce other biases. 
Conditioning on colliders may induce M-bias, although theoretical and empirical work suggests that such bias is typically small relative to the bias from failing to adjust for relevant confounders \citep{Liu2012Mbias, DingMiratrix2015}. 

Conditioning on instrumental variables can induce Z-bias. 
When unmeasured confounding persists, adjustment for instrumental variables may amplify omitted variable bias rather than reduce it \citep{Ding2017BiasAmplifier}. 
Even when all confounders are measured, including strong predictors of treatment that are not prognostic for the outcome may inflate variance \citep{Myers2011}.
Additional simulations do not indicate substantial degradation in heterogeneous treatment effect estimation arising from collider or instrumental variable adjustment. 
In our framework, the horseshoe prior regularises effect magnitudes in the outcome model, so covariates that primarily predict treatment but contribute little to outcome prediction tend to be shrunk strongly toward zero rather than explicitly excluded.
This proposed approach balances broad confounder adjustment with adaptive regularisation that mitigates the risks associated with high-dimensional covariate inclusion.

\section*{Acknowledgments}
The software implementation is available on CRAN: \url{https://cran.r-project.org/package=ShrinkageTrees}.
All data and replication code can be found at: \url{https://github.com/tijn-jacobs/ShrinkageTrees}.

\section*{Funding}
Tijn Jacobs and St\'ephanie van der Pas were funded by the European Union.
Views and opinions expressed are however those of the author(s) only and do not necessarily reflect those of the European Union or the European Research Council Executive Agency.
Neither the European Union nor the granting authority can be held responsible for them.
This work is supported by ERC grant BayCause, nr.\ 101074802.

\bibliographystyle{plainnat}
\bibliography{references}

\newpage
\appendix
\renewcommand{\thesection}{S.\arabic{section}}
\setcounter{section}{0}

\section*{Supplementary Materials}
\addcontentsline{toc}{section}{Supplementary Materials}

\section{Extra details to the Outer Gibbs sampler}

This section provides additional technical details on the outer Gibbs sampler described in Section~4.1 \citep{main_article}. 
A summary of the full sampling scheme is given in Algorithm~1. 
We elaborate on two key components: (i) the update of the error variance $\sigma^2$ based on its conjugate inverse-gamma posterior, and (ii) the optional use of the invariant treatment parametrisation introduced by \citet{hahn2020bayesian}, which requires additional Gibbs updates for treatment-specific scaling parameters. 
These details support a more complete understanding and implementation of the full posterior sampling procedure.

The noise is modelled as $\varepsilon_i \sim \mathcal{N}(0, \sigma^2)$, with the error variance $\sigma^2$ assigned a scaled inverse-chi-squared prior:
\begin{equation}
\sigma^2 \sim \text{Inv-}\chi^2(\nu, \psi),
\end{equation}
where $\nu$ denotes the degrees of freedom parameter and $\psi$ the prior scale. We follow \citet{BARToriginal}, setting $\nu = 3$ and choosing $\psi$ so that the prior is centred around a rough estimate of $\sigma^2$.

The variance of the error is sampled from its conditional distribution.
Recall that the prior is $\sigma^2 \sim \text{Inv-}\chi^2(\nu, \psi)$, which corresponds to an inverse-gamma distribution with shape parameter $\frac{1}{2}\nu$ and scale parameter $\frac{1}{2}\nu \psi$. 
By conjugacy, the posterior for $\sigma^2$ is inverse-gamma with updated shape parameter $\frac{\nu + n}{2}$ and scale parameter $\frac{1}{2}\bigl(\nu \psi + \sum_{i=1}^n r_i^2\bigr)$, where $r_i = \log{T^o_i} - \log{\widehat{T}_i}$ denotes the residual for individual $i$, computed using Equation~\eqref{model}.
Detailed derivations and further discussion can be found in \citet[Section 3.2 and 14.2]{Gelman2013BDA}.

We adopt the conventional binary treatment indicator $A$, coded as 1 for treated individuals and 0 for controls, to clearly communicate our model structure and causal assumptions. 
However, the proposed model can also accommodate the invariant parametrisation introduced by \citet{hahn2020bayesian}, which relies on a data-adaptive treatment coding scheme.  
Under the invariant parametrisation, the model is re-expressed as:
\begin{equation}
    \log T(a) = f(x, \hat{e}(x)) + b_a \cdot \tau(x) + \varepsilon,
\end{equation}
where $\varepsilon \sim \mathcal{N}(0, \sigma^2)$ and:
\begin{equation}
    b_1 \sim \mathcal{N}(0, 1/2) \quad \text{and} \quad b_0 \sim \mathcal{N}(0, 1/2).
\end{equation}
In this formulation, the CATE is given by:
\begin{equation}
    \text{CATE}(x) = (b_1 - b_0)\tau(x).
\end{equation}
To implement this approach, two additional update steps are included in the Gibbs sampler to sample $b_1$ and $b_0$. 
Conditional on the prognostic and treatment effect forests, and $\sigma$, these updates correspond to standard linear regression updates. 
Specifically, a linear model where the response is given by $\log T - f(x, \hat{e}(x))$ and the design matrix has two columns (without an intercept), defined as $(A\tau(x), (1-A)\tau(x))$. 
 \newpage
\section{Reversible jump MCMC for tree updates}\label{appendix:comp_details}

This section provides technical details on the reversible jump Metropolis--Hastings updates used to sample tree structures and their associated parameters in our model.
We update the tree structure $\mathcal{T}$ and step heights $\mathcal{H}$ jointly using the reversible jump framework \citep{ReversibleJump}.
We denote the current tree by $(\mathcal{T}, \mathcal{H})$ and the proposed tree by $(\mathcal{T}*, \mathcal{H}*)$.
We consider three reversible jump moves: \texttt{GROW}, \texttt{PRUNE}, and \texttt{CHANGE}, selected with fixed prior probabilities $P_\texttt{GROW}$, $P_\texttt{PRUNE}$, and $P_\texttt{CHANGE}$.
These moves enable exploration of the posterior by proposing local modifications to $(\mathcal{T}, \mathcal{H})$.

This section is organised as follows:
\begin{itemize}
    \item We begin with a brief recap of the model setup and prior specification.
    \item We then provide a detailed derivation of the acceptance ratio for the \texttt{GROW} move, which involves a change in model dimensionality.
    \item Finally, we explain how the acceptance ratios for the \texttt{PRUNE} and \texttt{CHANGE} moves follow from similar principles.
\end{itemize}

An overview of the steps involved in a single-tree reversible jump update is given in Algorithm~2 \citep{main_article}. 

\subsection{Recap of model setup and prior specification}

We model the log-transformed survival times as a noisy sum of regression trees:
\begin{equation}\label{model_appendix}
\log T_i = \sum_{j=1}^m g(X_i; \mathcal{T}_j, \mathcal{H}_j) + \varepsilon_i, \quad \varepsilon_i \sim \mathcal{N}(0, \sigma^2),
\end{equation}
where $m$ denotes the number of trees, $\mathcal{T}_j$ is the tree structure, and $\mathcal{H}_j$ is the set of step height parameters for the $j$-th tree.

We place independent priors on each pair $(\mathcal{T}_j, \mathcal{H}_j)$. The prior on the tree structure $\mathcal{T}$, originally proposed by \citet{BayesianCART}, is defined as follows:
\begin{enumerate}
    \item Start with a single root node.
    \item A leaf node at depth $d$ is split with probability:
    \begin{equation}\label{tree_prior}
        \rho_d := \frac{a}{(1 + d)^b},
    \end{equation}
    where $a \in (0, 1)$ and $b \geq 0$ are hyperparameters.
    \item The splitting variable is chosen uniformly at random from the available predictors.
    \item The split point is chosen uniformly from the observed values of the selected variable.
\end{enumerate}

The prior on the step height parameters $\mathcal{H}$, including any auxiliary parameters (e.g., local shrinkage parameters), is denoted by $p_\mathcal{H}(\mathcal{H} \mid \mathcal{T})$. Details can be found in Section~3.2 \citep{main_article}.

Posterior updates for $(\mathcal{T}, \mathcal{H})$ are based on the residuals derived from the current model fit. For the $j$-th tree, these are given by:
\begin{equation}\label{residuals_appendix}
\mathcal{R}_j := \log T^o - \sum_{J \ne j} g(X; \mathcal{T}_J, \mathcal{H}_J),
\end{equation}
where $T^o$ denotes the observed or imputed survival times after data augmentation.

\subsection{The \texttt{GROW} move}

When a \texttt{GROW} move is selected with probability $P_{\texttt{GROW}}$, the algorithm proposes to expand the current tree by splitting one of its terminal nodes (a leaf) into two new child nodes. 
This move increases the dimension of the parameter space and therefore requires a reversible jump step to ensure correct posterior exploration. 
The split is performed by selecting a valid splitting rule---that is, a covariate and a split point---which partitions the observations within the chosen parent node.

Consider a tree with three terminal nodes and corresponding step heights ${h_1, h_2, h_3}$, as illustrated on the left side of Figure~\ref{fig:GROWN_tree}. 
Suppose we select the third leaf node (associated with $h_3$) for a \texttt{GROW} move. 
A splitting rule is applied to this node, yielding two new child nodes. 
These new regions inherit a subset of the observations from the parent node, based on the splitting rule, and are each assigned their own proposed step heights, denoted $h_3'$ and $h_4'$. 
The resulting tree, shown on the right side of Figure~\ref{fig:GROWN_tree}, now contains four leaf nodes.

\begin{figure}[tb]
\centering
\begin{tikzpicture}[
    box/.style={rectangle, draw, rounded corners, align=center, minimum height=10mm, minimum width=16mm},
    leaf/.style={circle, draw, align=center, minimum size=10mm},
    line/.style={-latex},
    grow/.style={thick, ->, >=latex}
]

\begin{scope}
    \node [box] (root) {$x_1 < 0.7$};
    \node [box, below=1cm of root, xshift=-2cm] (left) {$x_2 < 0.6$};
    \node [leaf, below=1cm of root, xshift=2cm] (h1) {$h_1$};
    \node [leaf, below=1cm of left, xshift=-1cm] (h2) {$h_2$};
    \node [leaf, below=1cm of left, xshift=1cm] (h3) {$h_3$};

    \draw[line] (root) -- (left) node[midway, above left] {Yes};
    \draw[line] (root) -- (h1) node[midway, above right] {No};
    \draw[line] (left) -- (h2) node[midway, above left] {Yes};
    \draw[line] (left) -- (h3) node[midway, above right] {No};
\end{scope}

\draw[->, thick] (2.75, -2) -- (3.75, -2)  node[midway, sloped, above] {\texttt{GROW}}; 

\begin{scope}[xshift=7cm]
    \node [box] (root2) {$x_1 < 0.7$};
    \node [box, below=1cm of root2, xshift=-2cm] (left2) {$x_2 < 0.6$};
    \node [leaf, below=1cm of root2, xshift=2cm] (h1b) {$h_1$};
    \node [leaf, below=1cm of left2, xshift=-1cm] (h2b) {$h_2$};
    
    \node [box, below=1cm of left2, xshift=1cm] (split3) {$x_1 < 0.4$};
    \node [leaf, below=1cm of split3, xshift=-1cm] (h3p) {$h_3'$};
    \node [leaf, below=1cm of split3, xshift=1cm] (h4p) {$h_4'$};

    \draw[line] (root2) -- (left2) node[midway, above left] {Yes};
    \draw[line] (root2) -- (h1b) node[midway, above right] {No};
    \draw[line] (left2) -- (h2b) node[midway, above left] {Yes};
    \draw[line] (left2) -- (split3) node[midway, above right] {No};
    \draw[line] (split3) -- (h3p) node[midway, above left] {Yes};
    \draw[line] (split3) -- (h4p) node[midway, above right] {No};
\end{scope}

\end{tikzpicture}
\caption{Example of a \texttt{GROW} move on a decision tree: the left tree is expanded by splitting the leaf node corresponding to $h_3$. This yields two new leaves corresponding to step heights $h_3'$ and $h_4'$.}
\label{fig:GROWN_tree}
\end{figure}

The same transition can be viewed in the covariate space in Figure~\ref{fig:partition_grow}.
The left square represents the region before the split (corresponding to $h_3$) and the right square shows the refined partition into $h_3'$ and $h_4'$. 
This diagram offers geometric intuition for how the \texttt{GROW} move expands the parameter space by refining the partition of the covariate domain.

\begin{figure}[h]
    \centering
\begin{tikzpicture}[scale=3] 
    \draw[thick] (0, 0) rectangle (1, 1);
    
    \draw[thin] (0.7, 0) -- (0.7, 1) ;
    
    \draw[thin] (0, 0.6) -- (0.7, 0.6);
    
    \node[below] at (0.35, -0.1) {$x_1$};
    \node[left] at (-0.1, 0.25) {$x_2$};

    
    \node at (0.85, 0.5) {$h_1$};
    \node at (0.35, 0.3) {$h_2$};
    \node at (0.35, 0.8) {$h_3$};

    \draw[dashed] (0.85, 0.5) circle [radius=0.10];
    \draw[dashed] (0.35, 0.3) circle [radius=0.10];
    \draw[dashed] (0.35, 0.8) circle [radius=0.10];
    
    \begin{scope}[shift={(2.0, 0.0)}]
        \draw[thick] (0, 0.00) rectangle (1, 1);
    
        \draw[thin] (0.7, 0) -- (0.7, 1);
        \draw[thin] (0, 0.6) -- (0.7, 0.6);
        \draw[thin] (0.4, 0.6) -- (0.4, 1.0);
        
        \node at (0.85, 0.5) {$h_1$};
        \node at (0.35, 0.3) {$h_2$};
        \node at (0.55, 0.8) {$h_4'$};
        \node at (0.2, 0.8) {$h_3'$};
    
        \draw[dashed] (0.85, 0.5) circle [radius=0.10];
        \draw[dashed] (0.35, 0.3) circle [radius=0.10];
        \draw[dashed] (0.55, 0.8) circle [radius=0.10];
        \draw[dashed] (0.2, 0.8) circle [radius=0.10];

        \node[below] at (0.35, -0.1) {$x_1$};
        \node[left] at (-0.1, 0.25) {$x_2$};
    \end{scope}

    \draw[->, thick] (1.15, 0.5) -- (1.85, 0.5) node[midway, sloped, above] {\texttt{GROW}}; 
\end{tikzpicture}
\caption{Illustration of a \texttt{GROW} move: the left panel shows the original covariate partition, and the right panel shows the updated partition with one additional split and two new step heights.}
\label{fig:partition_grow}
\end{figure}

Let $\texttt{P}$ denote the parent node being split, which has associated parameters $(h_\texttt{P}, \theta_\texttt{P})$, where $\theta_\texttt{P}$ represents local auxiliary parameters. 
For example, under the horseshoe prior with auxiliary variables (see Equation~12, \citealp{main_article}), we have $\theta_\texttt{P} = (\lambda_\texttt{P}^2, \nu_\texttt{P})$. 
The \texttt{GROW} move proposes two new sets of parameters for the child nodes: $(h_\texttt{L}, \theta_\texttt{L})$ and $(h_\texttt{R}, \theta_\texttt{R})$, where $\texttt{L}$ and $\texttt{R}$ denote the left and right child nodes as determined by the splitting rule. 

These proposed values are drawn from pseudo Gibbs proposal distributions, as described in Section~4.2 \citep{main_article}. 
The proposals are informed by the current parameters within the tree and the observed data: new values for the step heights and local parameters are generated from conditional distributions influenced by those of the parent node. 
While this resembles a Gibbs sampler, it is not a true one — the parent node’s parameters cannot be directly sampled since the node is replaced by two new children. 
This is not problematic as the Metropolis-Hastings acceptance step guarantees that the Markov chain targets the correct posterior, thereby ensuring convergence to the stationary distribution. 
The overarching goal is to construct proposals that are both informed and efficient.

We now derive the reversible acceptance ratio $r_\texttt{GROW}$ for a \texttt{GROW} move. 
Note that the acceptance probability is given by $\min\{1, r_\texttt{GROW}\}$. 
The general framework for deriving the reversible jump acceptance ratio is outlined by \citet{ReversibleJump}. 
Following this approach, the acceptance ratio builds on the Metropolis--Hastings formulation, as applied to tree-based models by \citet{BARToriginal}, and incorporates both the proposal densities of the step heights and the determinant of the Jacobian associated with the parameter transformation. 
A comprehensive overview of how to derive acceptance ratios for non-reversible updates in standard BART is provided by \citet{Kapelner2016}.

We must satisfy the dimension-matching condition introduced by \citet{ReversibleJump} to perform a valid reversible jump move. 
Suppose we split a leaf node with associated parameters $(h_\texttt{P}, \theta_\texttt{P})$ into two child nodes with new parameters $(h_\texttt{L}, \theta_\texttt{L})$ and $(h_\texttt{R}, \theta_\texttt{R})$. 
We can interpret the parent node as having already been split, but with the child nodes initially inheriting the same values as the parent. 
This construction yields the same conditional posterior distribution and facilitates a reversible proposal structure. 
A schematic overview of this matching is depicted in Figure~\ref{fig:dim_match}. 
The Jacobian of the transformation is equal to one because the new parameters are randomly drawn using the pseudo Gibbs proposal mechanism.

\begin{figure}[h]
    \centering
\begin{tikzpicture}[scale=3] 
    
    
    \draw[thick] (0, 0) rectangle (1, 1);
    
    \draw[thin] (0.7, 0) -- (0.7, 1) ;
    \draw[thin] (0, 0.6) -- (0.7, 0.6);
    \draw[thin] (0.4, 0.6) -- (0.4, 1.0);
    
    \node[below] at (0.35, -0.1) {$x_1$};
    \node[left] at (-0.1, 0.25) {$x_2$};

    
    \node at (0.85, 0.5) {$h_1$};
    \node at (0.35, 0.3) {$h_2$};
    \node at (0.55, 0.8) {$h_3$};
    \node at (0.2, 0.8) {$h_3$};
        
    \draw[dashed] (0.85, 0.5) circle [radius=0.10];
    \draw[dashed] (0.35, 0.3) circle [radius=0.10];
    \draw[dashed] (0.55, 0.8) circle [radius=0.10];
    \draw[dashed] (0.2, 0.8) circle [radius=0.10];
        
    \begin{scope}[shift={(2.0, 0.0)}]
        \draw[thick] (0, 0.00) rectangle (1, 1);
    
        \draw[thin] (0.7, 0) -- (0.7, 1);
        \draw[thin] (0, 0.6) -- (0.7, 0.6);
        \draw[thin] (0.4, 0.6) -- (0.4, 1.0);
        
        \node at (0.85, 0.5) {$h_1$};
        \node at (0.35, 0.3) {$h_2$};
        \node at (0.55, 0.8) {$h_4'$};
        \node at (0.2, 0.8) {$h_3'$};
    
        \draw[dashed] (0.85, 0.5) circle [radius=0.10];
        \draw[dashed] (0.35, 0.3) circle [radius=0.10];
        \draw[dashed] (0.55, 0.8) circle [radius=0.10];
        \draw[dashed] (0.2, 0.8) circle [radius=0.10];

        \node[below] at (0.35, -0.1) {$x_1$};
        \node[left] at (-0.1, 0.25) {$x_2$};
    \end{scope}

    \draw[->, thick] (1.15, 0.5) -- (1.85, 0.5) node[midway, sloped, above] {\texttt{GROW}}; 
\end{tikzpicture}
\caption{Illustration of the dimension-matching argument used in the \texttt{GROW} move. The left domain represents the initial tree configuration with a step height $h_3$ duplicated across two child regions. The right domain shows a proposed configuration where these regions are assigned distinct step heights $h_3'$ and $h_4'$.}
\label{fig:dim_match}
\end{figure}

We derive the acceptance ratio $r_\texttt{GROW}$ for a \texttt{GROW} move. 
This ratio compares the probability of proposing and accepting a move to a larger tree versus reversing that move. 
It is given by:
\begin{equation}
    r_\texttt{GROW} = \frac{q\big((\T, \H) \to (\T_*, \H_*)\big)}{q\big((\T_*, \H_*) \to (\T, \H)\big)} \cdot \frac{p\big((\T, \H) \ | \ \mathcal{R}, \sigma^2\big)}{p\big((\T_*, \H_*) \ | \ \mathcal{R}, \sigma^2\big)},
\end{equation}
where $q(\cdot \to \cdot)$ denotes the transition probability between tree states, and $p(\cdot \mid \mathcal{R}, \sigma^2)$ is the conditional posterior distribution given the residuals and noise level.

By applying Bayes' rule to the posterior terms, we rewrite the ratio as:
\begin{equation}
\begin{split}
    r_\texttt{GROW} 
    &= \underbrace{\frac{q\big((\T, \H) \to (\T_*, \H_*)\big)}{q\big((\T_*, \H_*) \to (\T, \H)\big)}}_{\text{transition ratio}} 
    \cdot 
    \underbrace{\frac{\mathcal{L}\big(\mathcal{R} \mid \T, \H, \sigma^2\big)}{\mathcal{L}\big(\mathcal{R} \mid \T_*, \H_*, \sigma^2\big)}}_{\text{likelihood ratio}} 
    \cdot 
    \underbrace{\frac{p\big(\T, \H\big)}{p\big(\T_*, \H_*\big)}}_{\text{prior ratio}},
\end{split}
\end{equation}
where $\mathcal{L}$ is the likelihood of the residuals (see Equation~10, \citealp{main_article}), and $p(\T, \H)$ denotes the joint prior over tree structure $\T$ and step height parameters $\H$.

The likelihood only changes for observations in the node being split. For all other observations, the likelihood remains the same. As a result, the likelihood ratio simplifies to:
\begin{equation}
    \frac{\mathcal{L}\big(\mathcal{R} \mid \T, \H, \sigma^2\big)}{\mathcal{L}\big(\mathcal{R} \mid \T_*, \H_*, \sigma^2\big)} = \frac{\prod_{i \in \texttt{P}} \mathcal{N}(\mathcal{R}_i \mid h_\texttt{P}, \sigma^2)}{\prod_{i \in \texttt{L}} \mathcal{N}(\mathcal{R}_i \mid h_\texttt{L}, \sigma^2) \cdot \prod_{i \in \texttt{R}} \mathcal{N}(\mathcal{R}_i \mid h_\texttt{R}, \sigma^2)},
\end{equation}
where $\texttt{P}$ is the parent node being split, and $\texttt{L}$ and $\texttt{R}$ are the resulting left and right child nodes.
We write $i \in \texttt{L}$ to denote that observation $i$ falls into node $\texttt{L}$ under the tree partitioning.

The prior ratio decomposes into the ratio of the tree structure priors and the prior on the step heights:
\begin{equation}
\begin{split}
    \frac{p(\T, \H)}{p(\T_*, \H_*)} &= \frac{p_\mathcal{T}(\T)}{p_\mathcal{T}(\T_*)} \cdot \frac{p_\mathcal{H}(\H \mid \T)}{p_\mathcal{H}(\H_* \mid \T_*)} \\
    &= \frac{\rho_d(1-\rho_{d+1})^2}{1 - \rho_d} \cdot \frac{p(h_\texttt{P} \mid \theta_\texttt{P}) \, p(\theta_\texttt{P})}{p(h_\texttt{L} \mid \theta_\texttt{L}) \, p(\theta_\texttt{L}) \cdot p(h_\texttt{R} \mid \theta_\texttt{R}) \, p(\theta_\texttt{R})},
\end{split}
\end{equation}
where $d$ is the depth of the split node, and $\rho_d$ is the depth-dependent internal node probability from the tree prior (see Equation~\eqref{tree_prior}).

The transition ratio accounts for move probabilities, available node choices, and proposal densities:
\begin{equation}
    \frac{q\big((\T, \H) \to (\T_*, \H_*)\big)}{q\big((\T_*, \H_*) \to (\T, \H)\big)} 
    = \frac{P_\texttt{GROW}}{P_\texttt{PRUNE}} \cdot \frac{k^{-1}}{N_*^{-1}} \cdot \frac{q_\texttt{PRUNE}(h_\texttt{P}, \theta_\texttt{P})}{q_\texttt{GROW}(h_\texttt{L}, \theta_\texttt{L}) \cdot q_\texttt{GROW}(h_\texttt{R}, \theta_\texttt{R})},
\end{equation}
where $k$ is the number of leaf nodes in the current tree, $N_*$ is the number of eligible internal nodes (nog nodes) in the proposed tree, and $q$ refers to the proposal distributions.
We use the notation $q_\texttt{MOVE}$ to denote the proposal distribution used in each move, which is typically constructed as a pseudo Gibbs draw informed by the current state and residuals.

\subsection{The \texttt{PRUNE} and \texttt{CHANGE} moves}

In a \texttt{PRUNE} move, a \textit{nog} node (a parent node with exactly two children) is selected, and its children are removed. 
This corresponds to merging two adjacent subregions in the covariate space. 
The \texttt{PRUNE} move is the exact inverse of the \texttt{GROW} move. 
As a result, its reversible jump acceptance ratio $r_\texttt{PRUNE}$ is the reciprocal of the acceptance ratio for the corresponding \texttt{GROW} move:
\begin{equation}
    r_\texttt{PRUNE} = \frac{1}{r_\texttt{GROW}}.
\end{equation}

To propose parameters for the resulting leaf node, we must construct a reverse mapping from the child parameters. 
Several strategies are possible, such as averaging the child values or restoring the original parent values. 
We adopt the latter: the \texttt{PRUNE} move reverts to the parameter values of the parent node prior to the corresponding \texttt{GROW} move.

The \texttt{CHANGE} move modifies the splitting rule of an internal node while keeping the tree structure and parameter dimension fixed. 
Since the number of parameters remains unchanged, the reversible jump acceptance ratio $r_\texttt{CHANGE}$ includes a likelihood ratio, a proposal ratio, and a prior ratio for the step height parameters.

Suppose we propose new values $(h_{\texttt{L}*}, \theta_{\texttt{L}*})$ and $(h_{\texttt{R}*}, \theta_{\texttt{R}*})$ to replace the current parameters $(h_\texttt{L}, \theta_\texttt{L})$ and $(h_\texttt{R}, \theta_\texttt{R})$. The acceptance ratio becomes:
\begin{equation}
\begin{split}
    r_\texttt{CHANGE} 
    &= \frac{\prod_{i \in \texttt{L}} \mathcal{N}(\mathcal{R}_i \mid h_\texttt{L}, \sigma^2) \prod_{i \in \texttt{R}} \mathcal{N}(\mathcal{R}_i \mid h_\texttt{R}, \sigma^2)}{\prod_{i \in \texttt{L}*} \mathcal{N}(\mathcal{R}_i \mid h_{\texttt{L}*}, \sigma^2) \prod_{i \in \texttt{R}*} \mathcal{N}(\mathcal{R}_i \mid h_{\texttt{R}*}, \sigma^2)} \\
    &\quad \times \frac{q_\texttt{CHANGE}(h_{\texttt{L}*}, \theta_{\texttt{L}*}) \, q_\texttt{CHANGE}(h_{\texttt{R}*}, \theta_{\texttt{R}*})}{q_\texttt{CHANGE}(h_\texttt{L}, \theta_\texttt{L}) \, q_\texttt{CHANGE}(h_\texttt{R}, \theta_\texttt{R})} \\
    &\quad \times \frac{p(h_\texttt{L} \mid \theta_\texttt{L}) \, p(\theta_\texttt{L}) \, p(h_\texttt{R} \mid \theta_\texttt{R}) \, p(\theta_\texttt{R})}{p(h_{\texttt{L}*} \mid \theta_{\texttt{L}*}) \, p(\theta_{\texttt{L}*}) \, p(h_{\texttt{R}*} \mid \theta_{\texttt{R}*}) \, p(\theta_{\texttt{R}*})}.
\end{split}
\end{equation}

We draw the proposed parameters conditionally on the current ones. 
Although their interpretation may shift under the new split, the Metropolis--Hastings acceptance step accounts for this and preserves posterior correctness.

\subsection{Computational considerations}
The proposed sampler follows the standard BART back-fitting strategy and uses reversible jump MCMC (RJMCMC) updates to simultaneously update the tree-structure and leaf-node parameters.
The proposed tree-structure moves (GROW, PRUNE, and CHANGE) use the same structural proposal mechanism as in BART.
In contrast to standard BART, each proposed structural move also proposes corresponding step-height parameters for the respective leaves under the shrinkage prior.
These additional step-height proposals require a small number of extra random draws per newly created or modified leaf from conditional distributions (e.g.\ inverse-gamma or normal updates, depending on the reparametrisation).
The sampler then accepts or rejects the joint proposal for the tree structure and associated step heights using the RJMCMC acceptance probability.

Runtime scales primarily with the total number of trees in the ensemble.
The default implementation uses two forests of equal size, yielding 400 trees in total, whereas standard BART typically uses 200 trees.
Empirically, runtime increases approximately linearly with the number of trees.
For the PDAC analysis ($n=130$, $p=3029$), a single MCMC chain with 5000 burn-in and 5000 posterior draws required approximately 105 seconds.
All computations were performed on a MacBook Air with an Apple M3 chip and 16~GB of RAM, running macOS~26.2.

\begin{figure}[tb]
  \centering
  \includegraphics[width=\linewidth]{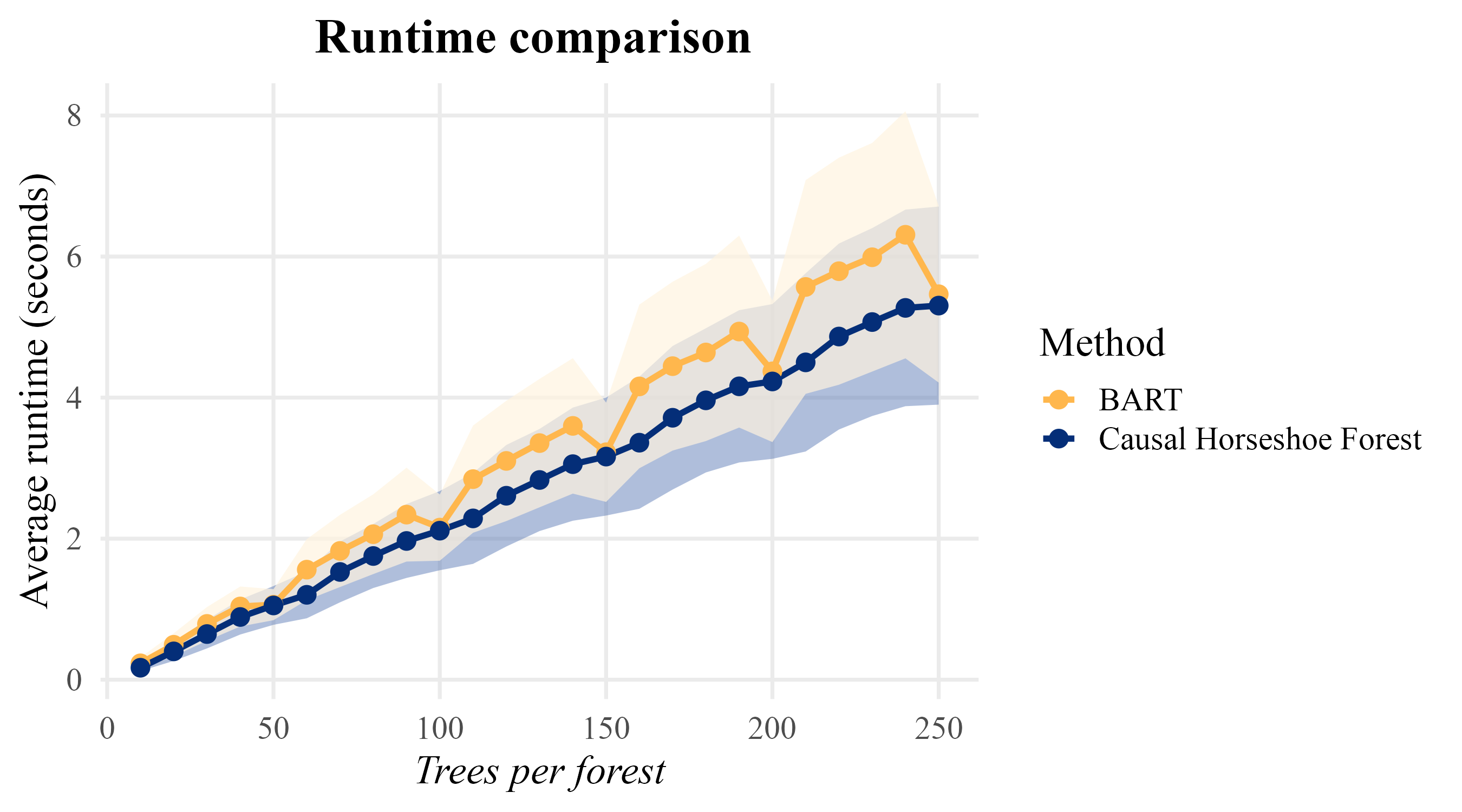}
  \caption{
    Average wall-clock runtime as a function of the number of trees per forest ($m$), averaged over five data-generating processes with 200 repetitions per setting. Shaded bands represent $\pm$ one standard deviation.
  }
  \label{fig:runtime}
\end{figure}

We conducted an empirical runtime comparison between Bayesian Additive Regression Trees (BART) \citep{bartRpackage} and the proposed Causal Horseshoe Forest (CHF) under uncensored data-generating processes.
Five distinct data-generating processes were considered, varying in sample size and covariate dimension.
Both methods were run under identical configurations, and wall-clock runtime was recorded in seconds.
This procedure was repeated 200 times per setting, resulting in a total of 1000 runs.
For the Causal Horseshoe Forest, we fixed the number of trees to $m$ per forest, yielding $2m$ trees in total across the treatment and control forests.
For BART, the ensemble size was fixed to $2m$ trees to ensure a comparable total number of trees across methods. Figure~\ref{fig:runtime} reports the average runtime as a function of $m$,
aggregated across data-generating processes and repetitions, with variability
summarised by one standard deviation.
Across the range of ensemble sizes considered, the two methods exhibit very
similar scaling behaviour.
In this implementation, CHF is marginally faster on average, which we attribute to implementation-level optimisations rather than fundamental algorithmic differences.
Overall, these results indicate that CHF achieves computational performance comparable to BART while offering additional modelling flexibility through explicit shrinkage priors.
 \newpage
\newpage
\section{Supplementary Tables for simulation results}\label{appendix:main_sim}

The tables below present full numerical results for the main simulation study.
These values correspond to the bar plots shown in Section~5.1 \citep{main_article} and are included here for reference and completeness.
They report RMSE, coverage, and average credible interval length for both CATE and ATE, across all methods and covariate dimensions.

\begin{table}[H]
\centering
\small
\renewcommand{\arraystretch}{0.9}
\begin{tabular}{clccc}
\toprule
$p$ & Method & RMSE & Coverage & Length \\
\midrule
100 & Causal Horseshoe Forest & 0.803 & 0.836 & 2.234 \\
  & AFT-BCF & 0.715 & 0.984 & 3.615 \\
  & AFT-Shrinkage BCF & 1.171 & 0.628 & 2.149 \\
  & AFT-BART & 1.036 & 0.291 & 0.874 \\
  & AFT-DART & 1.061 & 0.256 & 0.794 \\
\midrule
1000 & Causal Horseshoe Forest & 0.687 & 0.879 & 2.161 \\
  & AFT-BCF & 0.730 & 0.990 & 3.876 \\
  & AFT-Shrinkage BCF & 1.317 & 0.588 & 2.203 \\
  & AFT-BART & 1.150 & 0.220 & 0.824 \\
  & AFT-DART & 1.307 & 0.150 & 0.846 \\
\midrule
5000 & Causal Horseshoe Forest & 0.625 & 0.910 & 2.159 \\
  & AFT-BCF & 0.717 & 0.992 & 3.968 \\
  & AFT-Shrinkage BCF & 1.364 & 0.578 & 2.242 \\
  & AFT-BART & 1.075 & 0.172 & 0.617 \\
  & AFT-DART & 1.153 & 0.103 & 0.552 \\
\bottomrule
\end{tabular}
\caption{CATE performance for the linear scenario. Reported are RMSE, coverage, and average 95\% credible interval length across methods and dimensions.}
\label{tab:main_linear}
\end{table}

\begin{table}[H]
\centering
\small
\renewcommand{\arraystretch}{0.9}
\begin{tabular}{clccc}
\toprule
$p$ & Method & RMSE & Coverage & Length \\
\midrule
100 & Causal Horseshoe Forest & 0.564 & 0.896 & 1.814 \\
  & AFT-BCF & 0.484 & 0.998 & 2.863 \\
  & AFT-Shrinkage BCF & 0.493 & 0.838 & 1.421 \\
  & AFT-BART & 0.746 & 0.339 & 0.702 \\
  & AFT-DART & 0.776 & 0.433 & 0.952 \\
\midrule
1000 & Causal Horseshoe Forest & 0.563 & 0.881 & 1.764 \\
  & AFT-BCF & 0.575 & 0.997 & 3.202 \\
  & AFT-Shrinkage BCF & 0.792 & 0.701 & 1.689 \\
  & AFT-BART & 1.158 & 0.195 & 0.671 \\
  & AFT-DART & 1.277 & 0.275 & 1.070 \\
\midrule
5000 & Causal Horseshoe Forest & 0.482 & 0.926 & 1.727 \\
  & AFT-BCF & 0.576 & 0.999 & 3.605 \\
  & AFT-Shrinkage BCF & 1.111 & 0.614 & 1.989 \\
  & AFT-BART & 1.584 & 0.206 & 1.062 \\
  & AFT-DART & 1.851 & 0.236 & 1.455 \\
\bottomrule
\end{tabular}
\caption{CATE performance for the nonlinear scenario. Reported are RMSE, coverage, and average 95\% credible interval length across methods and dimensions.}
\label{tab:main_nonlinear}
\end{table}

 \newpage
\section{Additional high-dimensional regression study}

At the request of the referees, we conducted an additional simulation study in a simple regression setting without treatment assignment or censoring. 
The goal of this experiment is to allow for a clean comparison of shrinkage mechanisms within Bayesian tree ensembles. 
Accordingly, we restrict attention to Bayesian tree-based methods and do not include comparisons with non-Bayesian competitors such as random forests, boosting, or MARS. 
The focus is solely on how different regularisation strategies behave in high-dimensional non-linear regression.
The following simulations are inspired on the simulations performed by \citet{linero2018dirichlet} and \citet{LineroYang2018}.

Let $Y_i$ denote a continuous outcome and $X_i$ a $p$-dimensional covariate vector. 
We consider a single-forest tree ensemble model in which the outcome is represented as a noisy sum of regression trees:
\begin{equation}\label{bart_model}
Y_i = \sum_{j=1}^m g(X_i; \mathcal{T}_j, \mathcal{H}_j) + \varepsilon_i, 
\qquad 
\varepsilon_i \sim \mathcal{N}(0, \sigma^2),
\end{equation}
where each $\mathcal{T}_j$ is a binary decision tree defining recursive partitioning rules over the covariate space, and 
$\mathcal{H}_j = \{h_{j1}, \ldots, h_{jL_j}\}$ denotes the set of step heights associated with the $L_j$ terminal nodes of tree $\mathcal{T}_j$.
The function $g$ maps a covariate vector $X_i$ to the corresponding leaf-specific step height determined by the tree structure $\mathcal{T}_j$ and its associated parameters $\mathcal{H}_j$. 
That is, $g(X_i; \mathcal{T}_j, \mathcal{H}_j) = h_{jl}$ if $X_i$ falls into terminal node $l$ of tree $\mathcal{T}_j$.

We compare the following Bayesian tree ensemble methods, each of which induces a different form of regularisation:

\begin{itemize}

\item \textbf{BART} \citep{BARToriginal}.  
Regularisation is primarily imposed through the tree-structure prior, which penalises deep trees, and through Gaussian priors on the step heights $h_{jl}$. 
Shrinkage is global and homogeneous across leaves.

\item \textbf{DART} \citep{linero2018dirichlet}.  
Extends BART by placing a Dirichlet prior on splitting probabilities. 
This induces sparsity at the covariate-selection level by encouraging splits to concentrate on a subset of predictors.

\item \textbf{SoftBART} \citep{LineroYang2018}.  
Replaces hard splits with smooth gating functions and learns splitting weights adaptively. 
Regularisation is achieved through both smooth decision rules and hierarchical shrinkage on step heights.

\item \textbf{Horseshoe Forest} \citep{main_article}  
We use a single forest with a horseshoe global--local prior on the step heights.
Specifically, each step height is assigned a local shrinkage parameter while a tree-specific global shrinkage parameter controls overall regularisation. 
The default shrinkage level $k=0.1$ is used throughout.

\end{itemize}

For each individual $i = 1,\ldots,n$, we generate data according to
\begin{equation}
\begin{split}
    X_i &\sim \mathcal{U}[0,1]^p, \\
    \beta &\sim (1-s)\delta_{0_p} + s\,\mathcal{N}(0_p, I_p), \\
    f(x_i) &= 10 \sin(\pi x_{i1} x_{i2})
             + 20 (x_{i3} - 0.5)^2\\
             &\quad\quad + 10 x_{i4}
             + 5 x_{i5}
             + x_i^\top \beta, \\
    Y_i &\sim \mathcal{N}\big(f(x_i),\, \sigma^2\big),
\end{split}
\end{equation}
where $s = 0.05$ denotes the sparsity level. 
Each coefficient $\beta_j$ is drawn independently from a spike-and-slab distribution: it equals zero with probability $1-s$ and is sampled from a standard normal distribution with probability $s$.
The signal consists of a fixed non-linear component \citep{Friedman} depending on the first five covariates, and a sparse high-dimensional linear term $x_i^\top \beta$ with  sparsity level $s=0.05$. 
We set $\sigma^2$ such that $\mathrm{var}(Y)/\sigma^2 = 1.11\overline{1}$.

We first consider covariate dimensions $p \in \{500, 1000, 5000\}$ while fixing the sample size at $n = 100$. For each setting, we report the root mean squared error (RMSE), empirical coverage of the 95\% credible prediction intervals, and the average interval length.
Secondly, we investigate the high-dimensional behaviour in more detail by incrementally increasing $p$ from 50 to 5000. 
We plot the RMSE and empirical coverage of the posterior predictive intervals as functions of the covariate dimension $p$.

\subsection{Results}

\begin{figure}[tb]
\centering
\includegraphics[width=\linewidth]{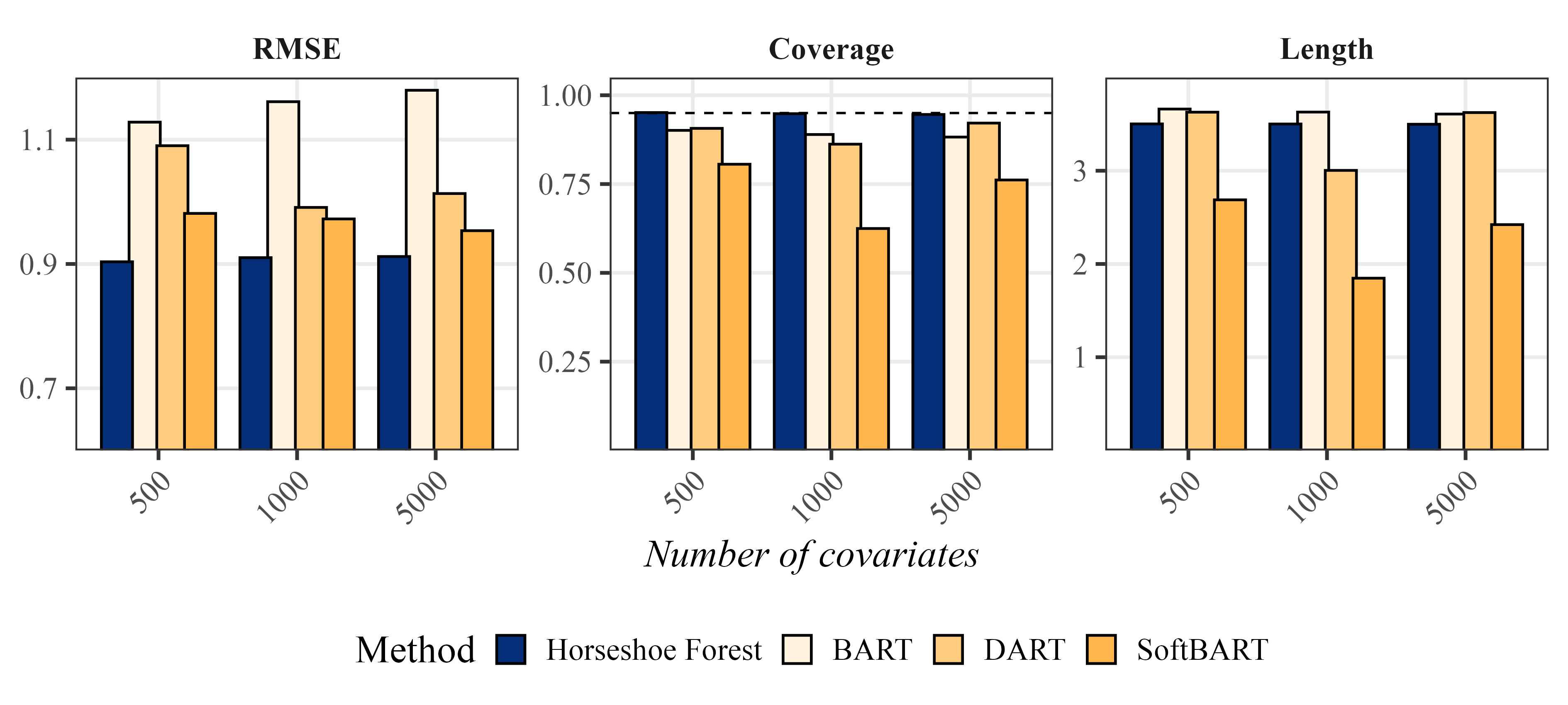}
\caption{
Performance comparison of Bayesian tree ensemble methods in a simple regression setting.
We report the root mean squared error (RMSE), empirical coverage of the 95\% posterior credible intervals, and the average credible interval length.
}
\label{fig:revision_simple_regression}
\end{figure}

The results for the settings $p \in \{500,1000,5000\}$ are shown in Figure~\ref{fig:revision_simple_regression}.
The Horseshoe Forest achieves the lowest RMSE across all three dimensions. 
Its predictive accuracy remains stable as $p$ increases. 
Standard BART yields consistently higher RMSE and does not improve with increasing $p$. DART and SoftBART reduce the error relative to BART.

The Horseshoe Forest attains empirical coverage close to the nominal $95\%$ level across all covariate dimensions while keeping interval lengths moderate. 
Standard BART achieves similar coverage but produces slightly wider intervals. 
DART shows some undercoverage, particularly in higher dimensions. 
SoftBART produces the shortest intervals but suffers from substantial undercoverage.

\begin{figure}[b!]
    \centering
    \includegraphics[width=\linewidth]{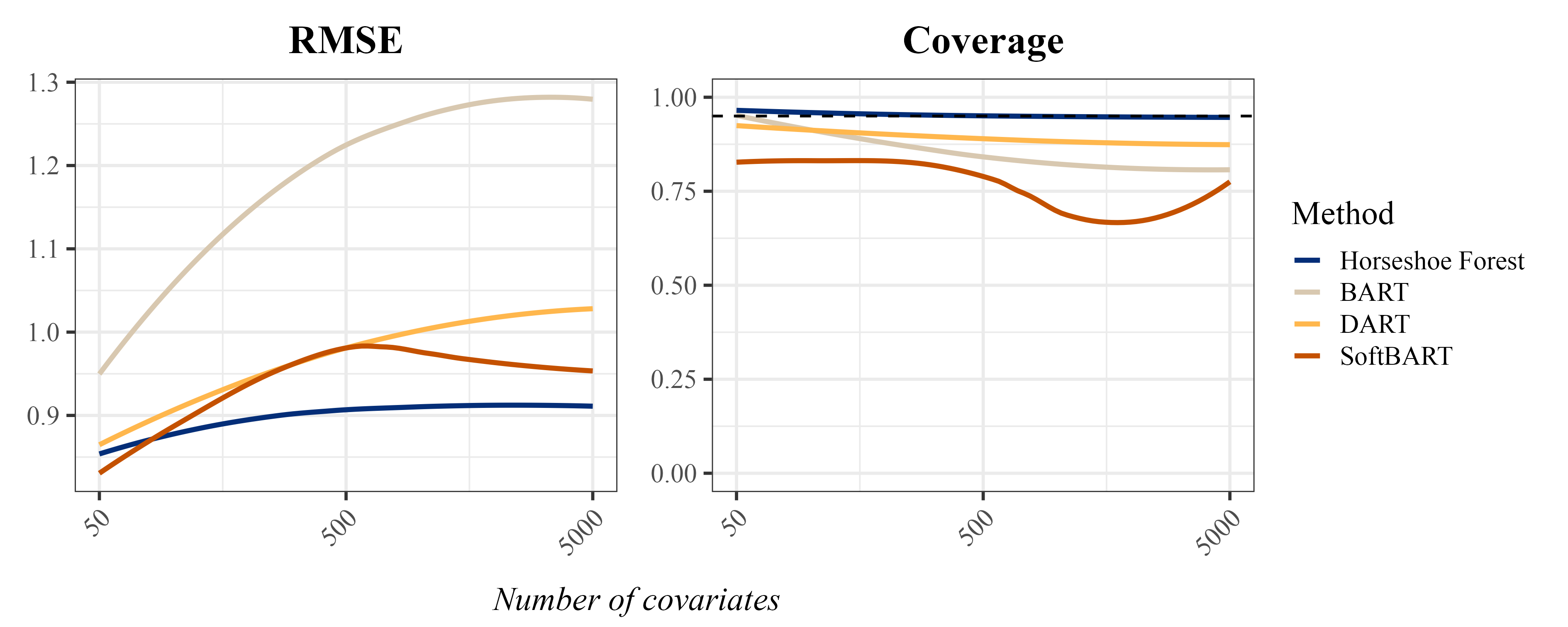}
    \caption{
    Performance comparison of Bayesian tree ensemble methods in a high-dimensional regression setting.
    The left panel shows the root mean squared error (RMSE) and the right panel shows the empirical coverage of 95\% posterior credible intervals as a function of the number of covariates $p$.
    }
    \label{fig:revision_simple_deeper}
\end{figure}

The results for $p$ varying from $50$ to $5000$ are shown in Figure~\ref{fig:revision_simple_deeper}. 
The Horseshoe Forest exhibits stable predictive accuracy across the entire range and achieves the lowest RMSE in moderate to high dimensions. 
Standard BART consistently shows the highest prediction error. 
DART improves upon standard BART and shows a noticeable gain in predictive accuracy once $p$ exceeds approximately $500$. 
SoftBART performs competitively in low dimensions, with relatively small prediction error when $p$ is close to $50$. 
As dimensionality increases, its RMSE rises in the moderate regime and then decreases slightly for larger $p$, leading to a mildly non-monotone pattern. 
The predictive performance of SoftBART remains between that of DART and the Horseshoe Forest in higher dimensions.

Coverage of the Horseshoe Forest remains close to the nominal $95\%$ level throughout. 
BART and DART shows moderate undercoverage at intermediate dimensions. 
SoftBART shows undercoverage 
SoftBART exhibits more pronounced undercoverage across most of the range. 
Its coverage declines further in the moderate-dimensional regime and only partially recovers for larger $p$, remaining below nominal. 

SoftBART displays mildly non-monotone behaviour in both RMSE and coverage as $p$ increases, with visible local fluctuations in the moderate-dimensional regime. 
Additional simulation runs confirm that this pattern is reproducible and not driven by Monte Carlo noise. 
While SoftBART remains computationally feasible across the considered dimensions, its runtime increases more markedly with $p$ than that of the Horseshoe Forest, reflecting the additional complexity of its soft splitting mechanism.

 \newpage
\section{Extra information on the PDAC analysis}
\label{appendix:clinical_covariates}
\raggedbottom

Although the main manuscript focuses on conditional treatment effects, we also report the average treatment effect (ATE) in the PDAC case study to facilitate comparison with existing clinical and methodological work. 
The ATE is defined as the population-level estimand:
\begin{equation}
\mathrm{ATE} := \E\!\left[\log T(1) - \log T(0)\right] = \E\left[\tau(x)\right],
\end{equation}
where the expectation is taken with respect to the covariate distribution in the target population.

We approximate this expectation by integrating the conditional average treatment effect over the empirical distribution of the observed covariates:
\begin{equation}
\widehat{\mathrm{ATE}}
= \frac{1}{n}\sum_{i=1}^n\widehat{\tau}(x_i).
\end{equation}
This plug-in estimator corresponds to averaging the posterior draws of the CATE over the observed covariate values and is standard in Bayesian tree-based causal models, including Bayesian causal
forests \citep{hahn2020bayesian}. 
The approach conditions on the observed covariates and therefore does not require specifying a probabilistic model for the covariate distribution.
This is also known as the MATE, or mixed average treatment effect \citep{Li2023BayesianCausal}.

Because the empirical covariate distribution is treated as fixed, this approximation does not explicitly propagate uncertainty in the covariate distribution itself.
More fully Bayesian alternatives, such as Bayesian bootstrap weighting of the covariate distribution, could be used to account for this additional source of uncertainty \citep{Oganisian2025Untangling}, but we do not
pursue these extensions here. 
Since our primary inferential focus is on heterogeneous treatment effects rather than average effects.

\begin{table}[H]
\centering
\label{tab:clinical_covariates}
\begin{tabular}{l p{9cm}} 
\toprule
\textbf{Variable} & \textbf{Description} \\
\midrule
\multicolumn{2}{l}{\textit{Demographic information}} \\
\texttt{age} & Age of the patient at diagnosis (in years) \\
\texttt{sex} & Biological sex (0 = female, 1 = male) \\
\addlinespace 
\multicolumn{2}{l}{\textit{Tumor characteristics}} \\
\texttt{grade} & Histological tumor grade \\
\texttt{tumor.cellularity} & Tumor cellularity as estimated by pathologist (numeric) \\
\texttt{tumor.purity} & Estimated tumor purity (0 = low, 1 = high) \\
\texttt{absolute.purity} & Tumor purity estimated using ABSOLUTE algorithm (numeric) \\
\addlinespace
\multicolumn{2}{l}{\textit{Molecular and expression features}} \\
\texttt{moffitt.cluster} & Molecular subtype (0 = classical, 1 = basal-like) \\
\texttt{meth.leukocyte.percent} & Estimated leukocyte fraction based on DNA methylation (numeric) \\
\texttt{meth.purity.mode} & Tumor purity from methylation model (numeric) \\
\addlinespace
\multicolumn{2}{l}{\textit{Staging and metastasis}} \\
\texttt{stage} & Nodal stage (0 = N0: no regional nodes, 1 = N1: nodal metastasis) \\
\texttt{lymph.nodes} & Number of lymph nodes examined (numeric) \\
\bottomrule
\end{tabular}
\caption{Overview of clinical covariates used in the PDAC data analysis.}
\end{table}

\begin{table}[H]
\centering
\label{tab:literature_genes}
\begin{tabular}{lll}
\toprule
\textbf{Stefanoudakis et al. (2024)} & \textbf{Cicenas et al. (2017)} & \textbf{Takai \& Yachida (2015)} \\
\midrule
\textit{KRAS}    & \textit{BRCA1}    & \textit{ARID1A}  \\
\textit{TP53}    & \textit{BRCA2}    & \textit{ATM}     \\
\textit{CDKN2A}  & \textit{PALB2}    & \textit{RNF43}   \\
\textit{SMAD4}   &                   & \textit{KDM6A}   \\
                 &                   & \textit{PBRM1}   \\
                 &                   & \textit{MLL3} \\
                 &                   & \textit{ARID2}   \\
                 &                   & \textit{SF3B1}   \\
                 &                   & \textit{BRAF}    \\
                 &                   & \textit{NRG1}    \\
                 &                   & \textit{RET}     \\
\bottomrule
\end{tabular}
\caption{Genes identified as drivers in pancreatic ductal adenocarcinoma (PDAC) according to the literature.}
\end{table}

\begin{figure}[ht]
    \centering
    \includegraphics[width=0.9\linewidth]{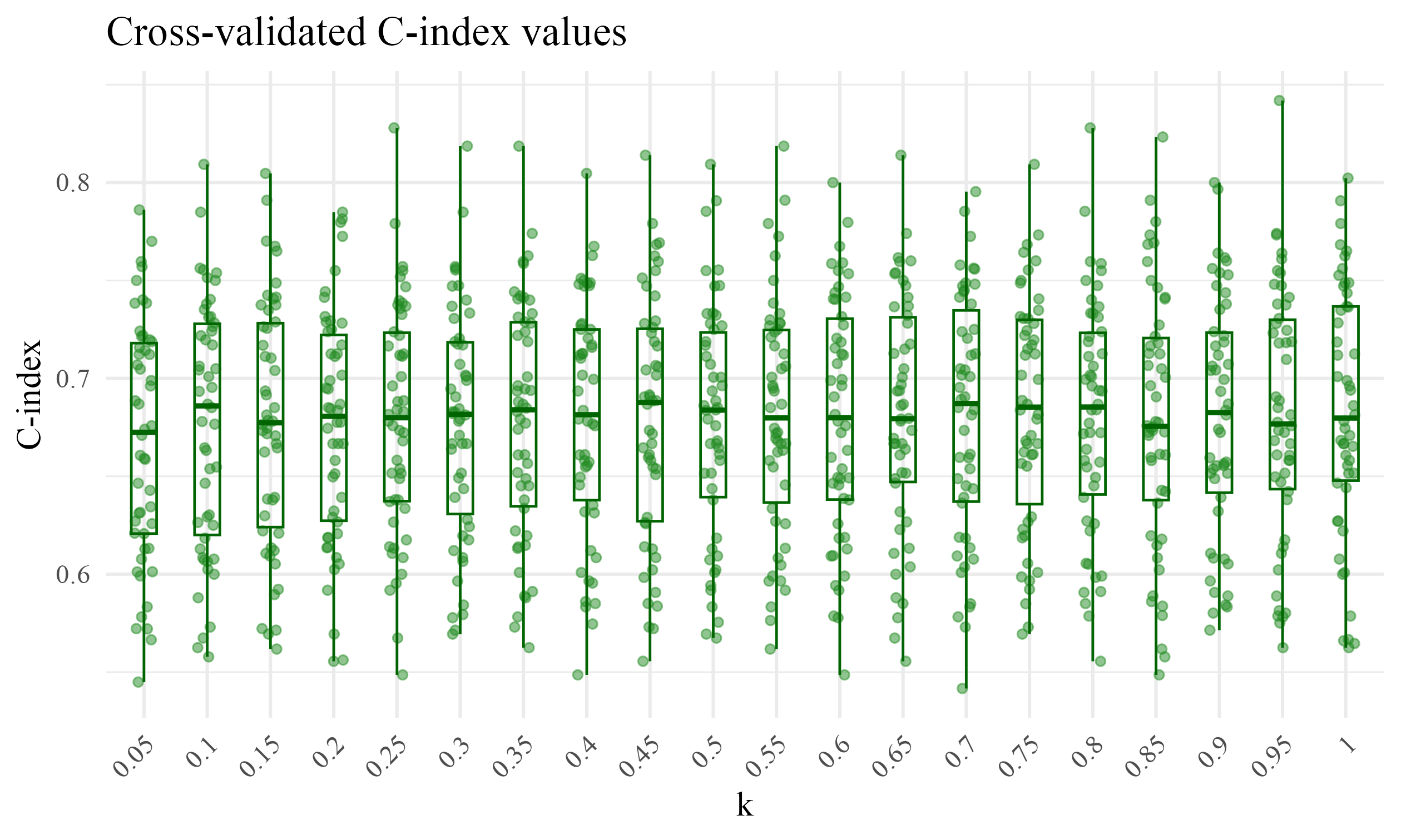}
    \caption{Stratified $10 \times 5$ cross-validated C-index values across different shrinkage values $k$.}
    \label{fig:cindex_k}
\end{figure}

\begin{figure}[H]
    \centering
    \includegraphics[width=\linewidth]{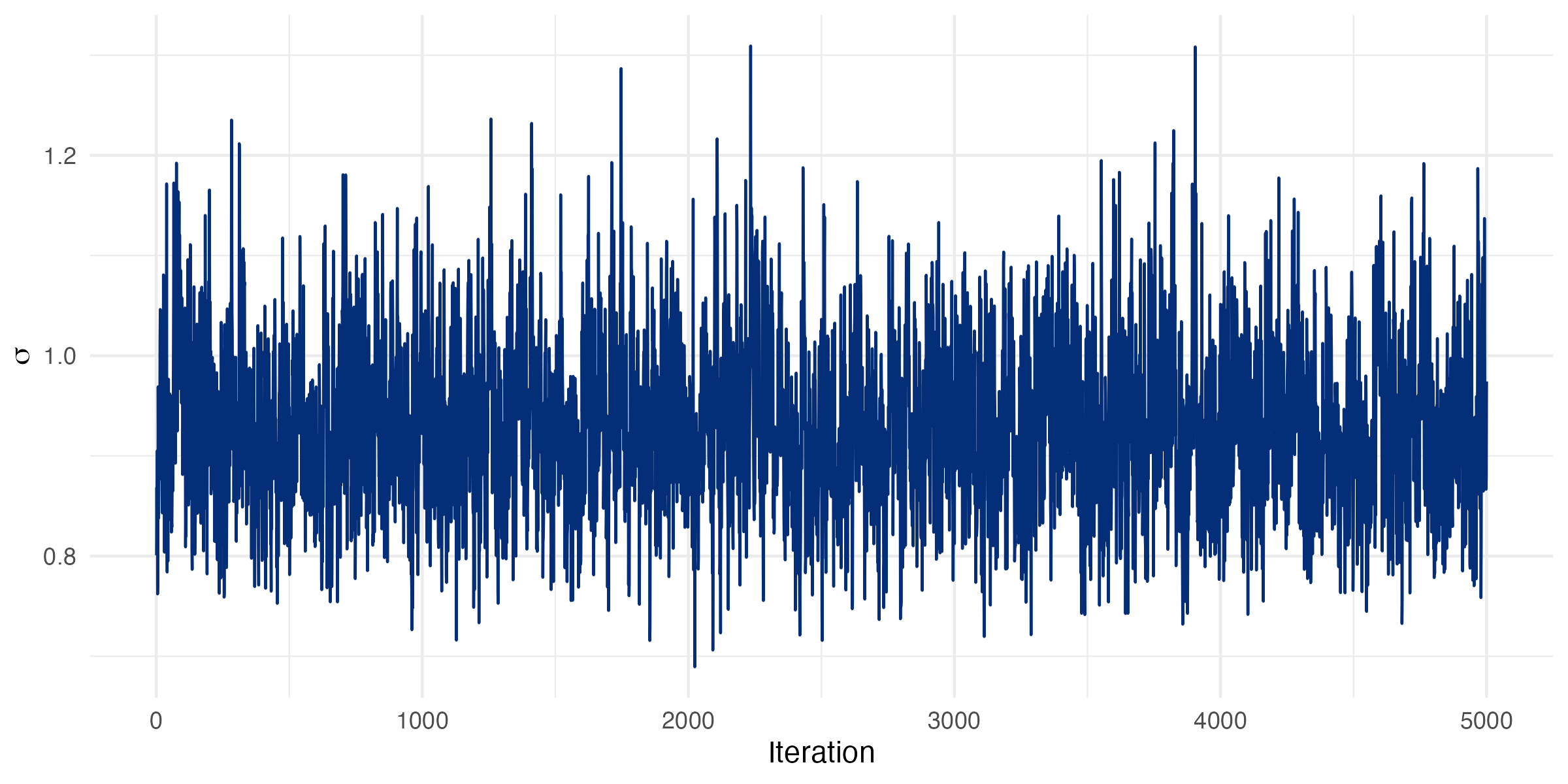}
    \caption{Traceplot of the posterior samples of $\sigma$, used to assess convergence of the Causal Horseshoe Forest sampler.}
    \label{fig:traceplot_sigma}
\end{figure}

\begin{figure}[H]
    \centering
    \includegraphics[width=0.9\linewidth]{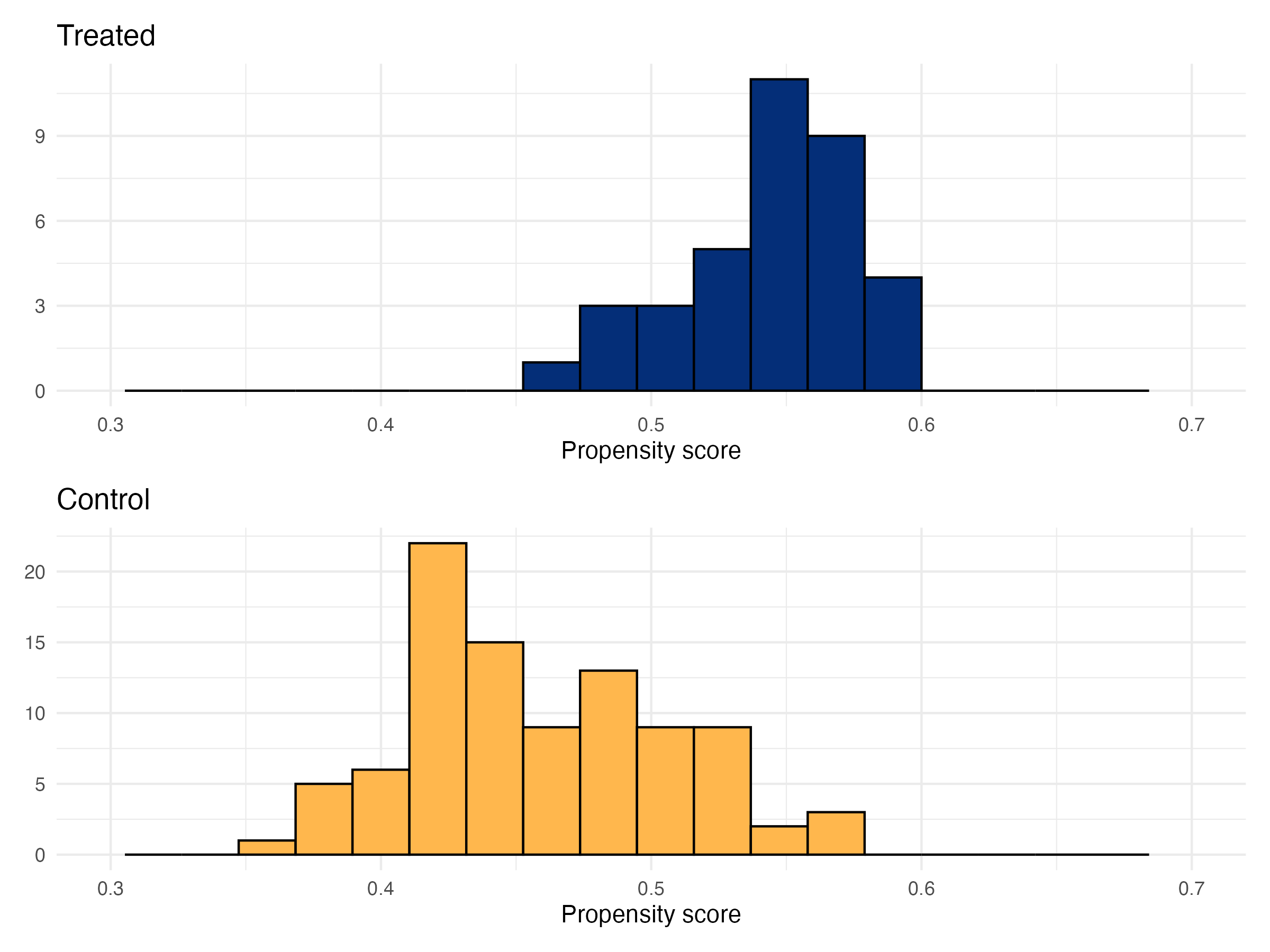}
    \caption{Histograms of estimated propensity scores for treated and control patients, showing the overlap in estimated treatment probabilities.}
    \label{fig:propensity_histograms}
\end{figure}

 \newpage
\section{Exploratory landmark analysis of PDAC data}\label{appendix:landmark}

\begin{figure}[b!]
    \centering
    \includegraphics[width=0.7\linewidth]{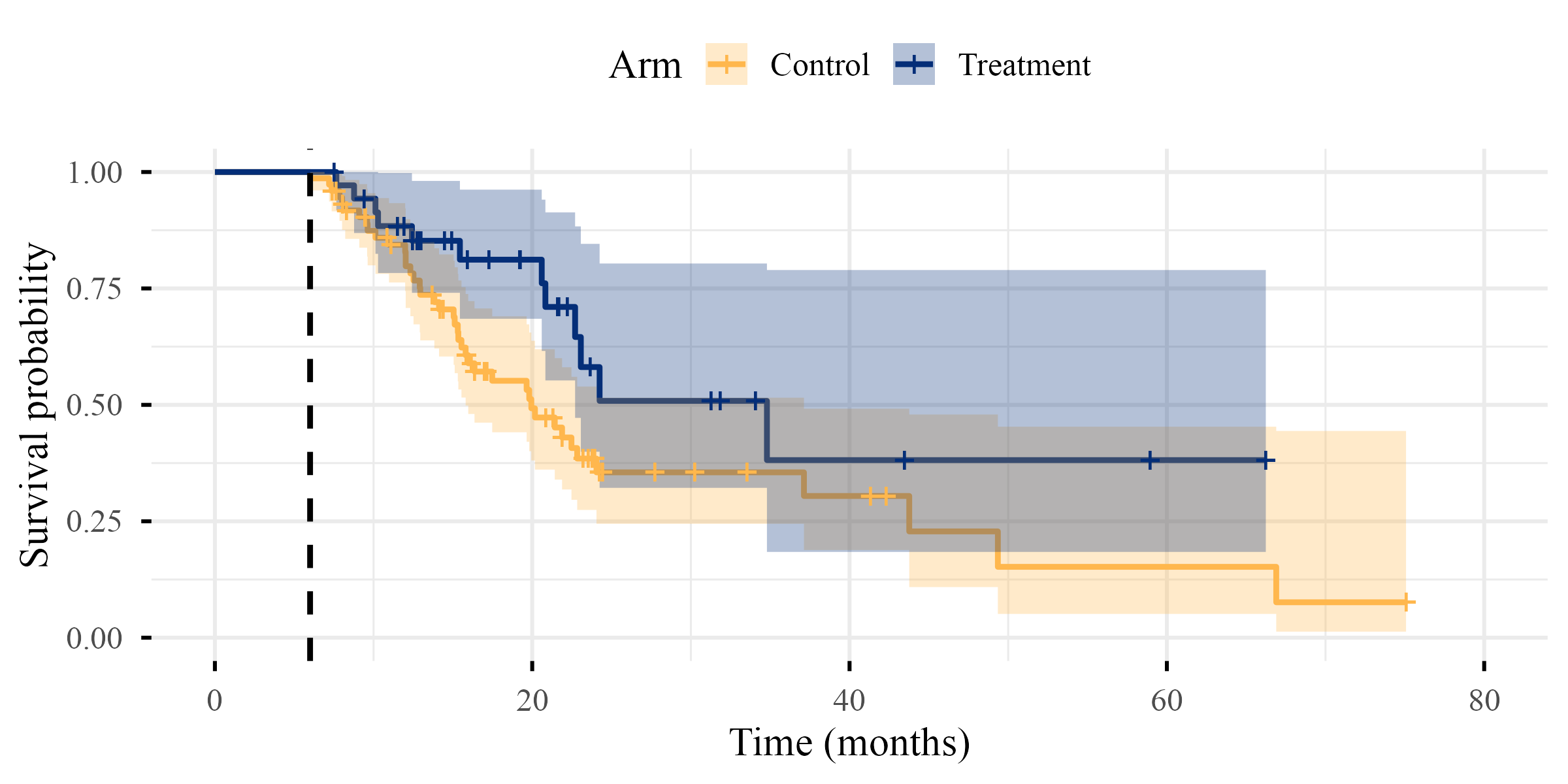}
    \caption{Kaplan--Meier survival curves for patients with PDAC after applying the 6-month landmark analysis. The vertical dashed line indicates the landmark time.}
    \label{fig:PDAC_KaplanMeier_Landmark}
\end{figure}

We conducted a preliminary landmark analysis on the PDAC data to explore the potential impact of immortal time bias.  
Two important caveats must be emphasised. 
First, we do not have detailed information on the exact timing of adjuvant radiation therapy initiation. 
We must therefore assume that at the chosen landmark time, patients classified as controls would not have subsequently received the treatment. 
Second, we applied our original model and estimands directly to the landmark cohort without formal adaptation to a rigorous landmark analysis framework. 
The results of this analysis should be interpreted with caution and viewed primarily as an exploratory sensitivity analysis, motivating further research rather than providing definitive causal conclusions.

The landmarking approach aims to mitigate immortal time bias by conditioning on survival up to a fixed time point.
It ensures that all included patients are comparably at risk beyond the landmark time. 
This strategy has been advocated in the survival analysis literature as a pragmatic way to reduce immortal time bias \citep{vanHouwelingen2007Dynamic}. 
However, fully addressing this bias requires explicit modelling of time-dependent treatment assignment and joint survival-treatment processes. 
This was not possible with the available data.

We set the landmark time at 6 months. 
Patients not at risk at this time point were excluded.
This results in a sample of $n = 110$ patients, with 36 in the treatment group and 74 in the control group (based on the original sample used in the main analysis). 
The corresponding Kaplan--Meier survival curves are shown in Figure~\ref{fig:PDAC_KaplanMeier_Landmark}. 
We then re-estimated the causal effects using the Causal Horseshoe Forest approach described earlier.

\begin{figure}[tb]
    \centering
    \begin{minipage}[ht]{0.48\linewidth}
        \centering
        \includegraphics[width=\linewidth]{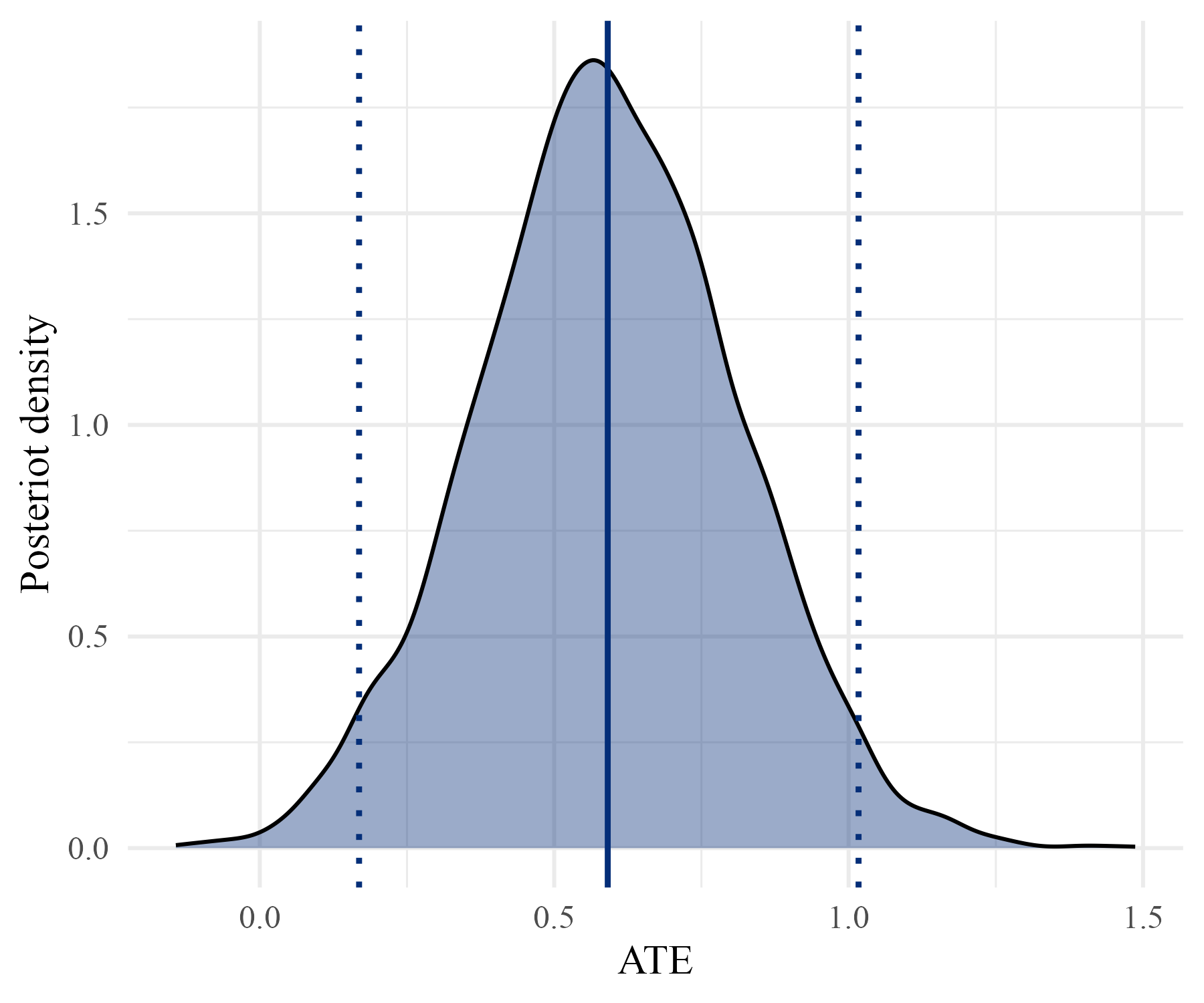}
        \caption{Posterior distribution of the ATE of adjuvant radiotherapy on overall survival in patients with PDAC after 6-month landmarking. The solid vertical line indicates the posterior mean, and dotted lines show the 95\% credible interval.}
        \label{fig:posterior_ATE_Landmark}
    \end{minipage}
    \hfill
    \begin{minipage}[ht]{0.48\linewidth}
        \centering
        \includegraphics[width=\linewidth]{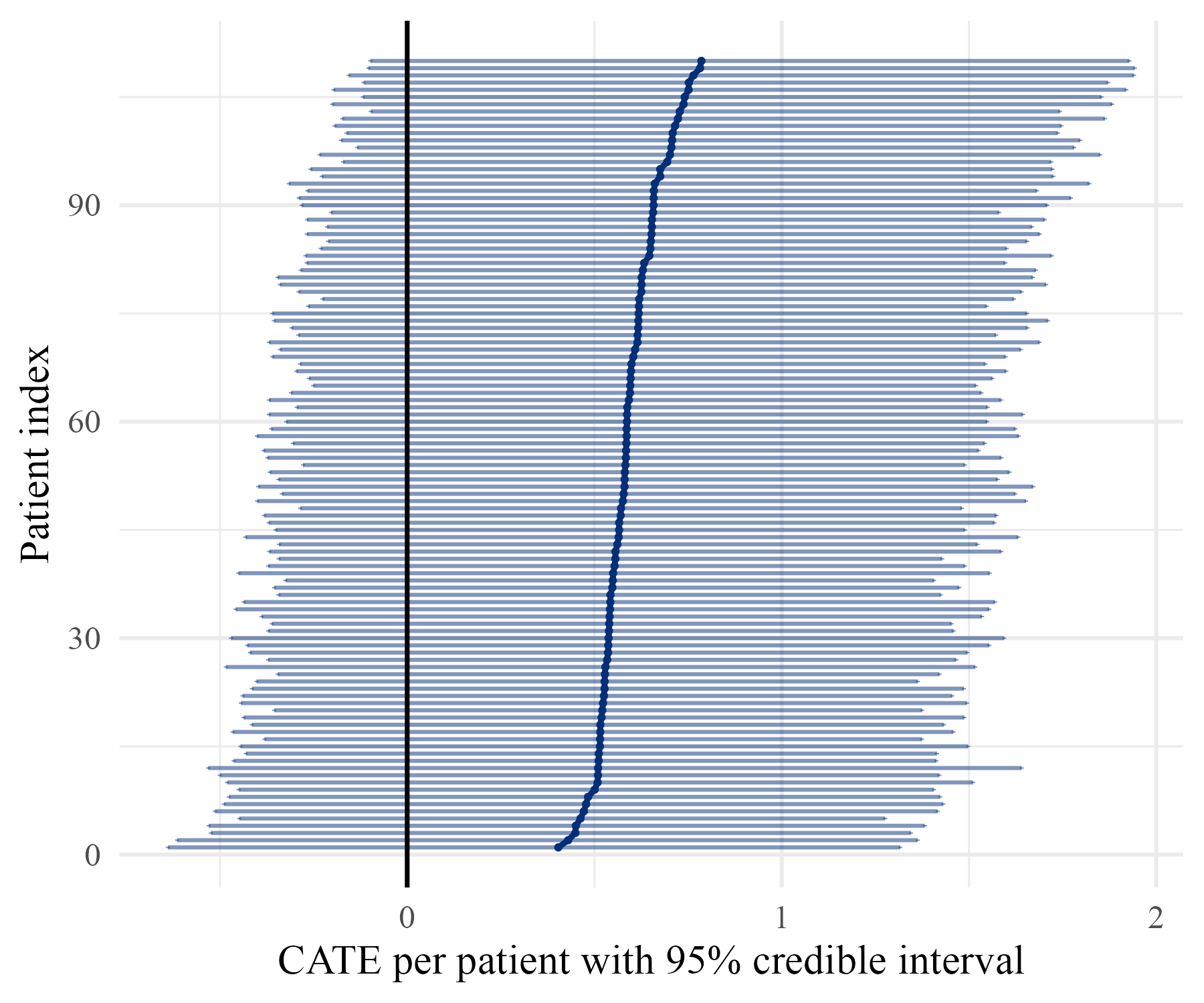}
        \caption{Posterior means and 95\% credible intervals for individual CATEs under adjuvant radiotherapy in patients with PDAC after 6-month landmarking. Patients are sorted by estimated CATE.}
        \label{fig:cate_plot_Landmark}
    \end{minipage}
\end{figure}

The model achieved a concordance index of $0.75$. 
The posterior mean ATE was estimated at approximately $0.55$, with a 95\% credible interval of ($0.10$, $0.98$), as shown in Figure~\ref{fig:posterior_ATE_Landmark}. 
Posterior distributions of the individual conditional average treatment effects are displayed in Figure~\ref{fig:cate_plot_Landmark}. 
The estimated CATEs suggest a relatively homogeneous treatment effect across patients, although the wide credible intervals highlight the uncertainty due to the smaller sample size and reduced number of events after landmarking.

The landmark analysis suggests a positive average treatment effect after accounting for potential immortal time bias. 
The estimated effect is smaller than in the main analysis.
This suggests that immortal time bias may have inflated the treatment effect observed previously.
The interpretation of this analysis crucially depends on the assumption that all patients would have had the opportunity to receive and complete adjuvant radiation therapy within one year after surgery.  
Future work incorporating explicit time-dependent treatment modelling is warranted to rigorously evaluate this potential bias.

\end{document}